\documentclass[manuscript,screen]{acmart}

\usepackage[braket,qm]{qcircuit}

\mathchardef\mathcomma=\mathcode`,

\AtBeginDocument{%
  \providecommand\BibTeX{{%
    \normalfont B\kern-0.5em{\scshape i\kern-0.25em b}\kern-0.8em\TeX}}}

\begin{document}

\title{Demonstration of a Hardware-Independent Toolkit for Automated Quantum Subcircuit Synthesis}

\author{Elena R. Henderson}
\email{erhenderson@smu.edu}
\orcid{0000-0003-4278-7617}
\author{Jessie M. Henderson}
\email{hendersonj@smu.edu}
\orcid{0000-0003-2848-8653}
\author{Aviraj Sinha}
\email{avirajs@smu.edu}
\orcid{0000-0002-5280-1585}
\author{Eric C. Larson}
\email{eclarson@smu.edu}
\orcid{0000-0001-6040-868X}
\author{Mitchell A. Thornton}
\email{mitch@smu.edu}
\orcid{0000-0003-3559-9511}
\affiliation{%
  \institution{Darwin Deason Institute for Cyber Security, Southern Methodist University}
  \streetaddress{6425 Boaz Lane Dallas}
  \city{Dallas}
  \state{Texas}
  \country{USA}
  \postcode{75205}}

\renewcommand{\shortauthors}{ERH, JMH, AS, ECL, MAT}

\begin{abstract}
The quantum computer has become contemporary reality, with the first two-qubit machine of mere decades ago transforming into cloud-accessible devices with tens, hundreds, or---in a few cases---even thousands of qubits.
While such hardware is noisy and still relatively small, the increasing number of operable qubits raises another challenge: how to develop the now-sizeable quantum circuits executable on these machines.
Preparing circuits manually for specifications of any meaningful size is at best tedious and at worst impossible, creating a need for automation.
This article describes an automated quantum-software toolkit for synthesis, compilation, and optimization, which transforms classically-specified, irreversible functions into both technology-independent and technology-dependent quantum circuits.
We also describe and analyze the toolkit's application to three situations---quantum read-only memories, quantum random number generators, and quantum oracles---and illustrate the toolkit's start-to-finish features, from the input of classical functions to the output of technology-dependent quantum circuits.
Furthermore, we illustrate how the toolkit enables research beyond circuit synthesis, including comparison of synthesis and optimization methods and deeper understanding of even well-studied quantum algorithms. 
As quantum hardware continues to develop, such quantum circuit toolkits will play a critical role in realizing its potential.
\end{abstract}


\maketitle

\section{Introduction}\label{sec:introduction}
The theory of quantum computers dates back to nearly the beginnings of classical computing.
Foundational concepts and the mathematics thereof were established contemporaneously with the first classical computer implementations \cite{neumann55}, and by the 1980s, physicists and computer scientists considered how to leverage---and not simply contend with---the quantum effects that arise when machine components become extremely small \cite{manin07,benioff80,feynman02,wiesner83,deustch85}.
Decades of subsequent work, including development of two of the most celebrated quantum algorithms to date \cite{grover96,shor94}, led to the first implementation of a quantum computer having two quantum bits---or qubits---comprised of ionized molecules controlled by a magnetic field \cite{chuang98}.
Today, production of quantum computers is a commercial enterprise using diverse approaches, from trapped-ion processors to photonics-based computers to increasingly popular superconducting-qubit machines \cite{tqd20}.

In every variant, the number of qubits remains quite limited, but is growing: quantum annealers have offered thousands of qubits for some time \cite{boothby19}, and the first one-thousand-qubit gate-based quantum machines were announced just last year \cite{computing23, castelvecchi23}.
As available hardware scales, so too does the size of circuits that can be run to completion, and quantum computers have begun to achieve results that are not obtainable on classical machines \cite{arute19,kim23}.
Consequently, a variety of classical software tools have been developed to address quantum circuit design and optimization \cite{serrano22}.
While many provide sophisticated optimization approaches for existing quantum circuits, there remains a gap: such tools require manual conversion into the first representation of a quantum circuit, whether that be implemented as quantum gates or represented by linear algebraic structures that can be transformed into gates.

Yet designing quantum circuits is an arduous task, as the fundamentally different physics and mathematics of quantum theory necessitate a tedious conversion from classical algorithms and data structures into their quantum counterparts.
The task is complicated further by the fact that quantum hardware still exists in the noisy, intermediate-scale (NISQ) era \cite{preskill18}, so quantum algorithm developers must work at the gate level to optimize for metrics including gate and qubit count.
Thus, the need for automated quantum circuit design tools now parallels that for classical electronic design automation (EDA) tools: it is either tedious or impossible to design large circuits manually, particularly when optimizing repeated or otherwise important subcircuits within larger specifications.
Furthermore, by automating synthesis and optimization, developer time can focus on higher-level algorithmic considerations, thereby streamlining the design process.

This need for a `quantum EDA toolkit' is heightened by the special importance that subcircuits have in many quantum algorithms.
First, quantum algorithms often contain subroutines that---when implemented as subcircuits---must be designed and optimized for the context of every particular problem.
For example, Grover's Search Algorithm includes a subcircuit termed an oracle, which has a single qualitative purpose, but whose particular implementation changes for both every database to be searched and for every search query \cite{grover96,henderson23}.
Second, many quantum algorithms require subcircuits for even more functionality than do their classical counterparts.
For example, when processing datasets using quantum machine learning (QML), the data must be provided using subcircuits termed `state generation circuits,' and the gates comprising these subcircuits are different for every unique dataset \cite{sinha22}.
Third, generating any subcircuit often requires not only determining the appropriate gate and qubit structure, but also substantial preprocessing of the classically-specified function prior to quantum circuit synthesis \cite{sinha22,sinha23,henderson23}.
Fourth and finally, subcircuits of varying types are often related, meaning that automated tools for synthesizing---or optimizing---one type of subcircuit can be leveraged for multiple applications \cite{sinha22,henderson23}.

In this article, we thus introduce a toolkit---\textit{MustangQ}---for automating quantum subcircuit design and optimization, and we proceed as follows.
Section \ref{sec:background} provides brief background on quantum computing and three types of quantum subcircuit.
Then, Section \ref{sec:MustangQ} introduces \textit{MustangQ}'s functionality, from its ability to preprocess classical function specifications to its available synthesis methods.
In, Section \ref{sec:MustangQ_subcircuits}, we illustrate our toolkit's ability to contribute to meaningful research by presenting automated synthesis of three subcircuit types, explaining how such circuit generation supports more robust analysis, and discussing the research conclusions that thereby follow.
Finally, we conclude by considering avenues for further research and development.

Before moving further, it is worth clarifying the scope of this paper.
This work is a comprehensive introduction to our toolkit and preliminary theoretical analysis, but is not designed to benchmark \textit{MustangQ}'s circuits on actual quantum devices, nor to assess related inquiries of noise, decoherence, or error mitigation.
We chose to omit such discussion for four reasons.
First and foremost, we face a practical difficulty of obtaining sufficient access to enough hardware to benchmark the three types of subcircuits we consider.
Second, one of the primary goals of \textit{MustangQ} is to design circuits that are not feasible for limited-qubit hardware, but that are expected to be implementable on future machines.
Third, even for the \textit{MustangQ} subcircuits that are small enough to be run on today's machines, hardware implementation and error mitigation are substantial enough topics to merit individual studies for each of the subcircuit types considered here.
(For example, the theory and implementation of subcircuits similar to those in Section \ref{sub:MustangQ_QRNGs} have been studied separately; see Refs. \cite{li21} and \cite{orts23}.)
Fourth and finally, we felt that the length of this paper was such that it would not benefit readers to further expand it.
All of these reasons notwithstanding, questions of hardware implementation are of the utmost import, and we do plan to address them in future work, especially for the subcircuits of Section \ref{sub:MustangQ_QRNGs}, which are sized favorably for contemporary hardware.

\section{Background}\label{sec:background}
\subsection{A Brief Introduction to Quantum Computing}\label{sub:intro_to_quantum}
Quantum computing leverages the fundamentally different physics---and thus the fundamentally different mathematics---of quantum mechanics to accomplish tasks that a classical computer cannot complete in any reasonable amount of time \cite{nielsen11}.
And yet, while the quantum trifecta of superposition, entanglement, and tunneling thus offers great theoretical promise, selecting applicable problems and then designing algorithms for them is anything but trivial.
Once designed, the mathematics of such algorithms must be realized as a quantum circuit, which---in the gate-based paradigm---represents the algorithm as a series of low-level operations (or, ``gates'') that are applied to qubits \cite{feynman02,deustch85,deustch89,divincenzo00,nielsen11}.

While the structures of many quantum circuits are well understood, several quantum algorithms contain subroutines that retain their function, but have differing circuit implementations, depending upon the situation.
When implemented as circuits, we term these quantum subroutines `unpatterned subcircuits.'
For example, several algorithms---including Grover's search algorithm---use subroutines termed quantum oracles, which have the function of `marking' states that satisfy a user-specified search query, while leaving non-solution states unmarked.
Oracles are thus blackbox functions that mark states according to some mathematical function, and this means quantum algorithms can be developed `around' such oracles, allowing for consideration of the overall process without requiring synthesis of a particular oracle for a particular contextual case.
However, when implementing such an algorithm in a quantum circuit for a particular situation, we cannot leave the oracle subroutine as an unspecified subcircuit; instead, we must specify its marking functionality using particular quantum gates.
Consequently, a quantum programmer designing circuits to execute Grover's search must implement different oracle subcircuits not only for every different database to be searched, but also for every different search query within a single database.
Thus, not only can such unpatterned subcircuit synthesis be extremely tedious, but it can also be impossible to complete manually for sizeable circuits, particularly when a developer wants to optimize qubit and gate count.

Before describing our toolkit for automated production of such subcircuits, we briefly introduce each of the three classes of quantum subcircuit to which we will apply it: quantum read-only memories (QROM), quantum random number generators (QRNG), and quantum oracles.
While these have distinct objectives, they all serve as bridges for transforming classical information to quantum algorithms represented as quantum circuits.
Specifically, QROM generates quantum representations of addressable data, which can be accessed in subsets (analogous to a classical read-only memory) and which can be processed by QML.
QRNGs represent probability distributions specified by probability mass functions (PMFs).
And quantum oracles represent classically-specified functions that mark certain quantum states, according to problem-specific information.
While these subcircuit classes can have different specifications, they often complement each other; for example, a QROM might load a dataset that could then be processed with an algorithm that filters states using an oracle.
Furthermore, QROMs, QRNGs, and oracles can all use quantum data encoding methods to represent relevant information using qubits.
In fact, QROMs can be understood as a subclass of oracles, as they `mark' data associated with specified addresses.
The subsequent sections briefly introduce each subcircuit type and some of its most salient features.
Table \ref{table:subcircuit_summary} summarizes mathematical representations of these subcircuits.

\subsection{Quantum Read-only Memory Subcircuits}\label{sub:quantum_memory}
Quantum circuits can represent analogs of classical memories---such as random access memory (RAM)---which can then be processed with algorithms such as those in the field of QML.
However, unlike for classical data processing, the input data for quantum processing applications must be generated, by evolving qubits from basis states into states representing the information to be processed \cite{schuld18,ventura00,phalak22}.
This evolution occurs via the application of quantum gates, and thus requires synthesizing state generation subcircuits that realize the desired input data \cite{travaglione01}. 

Quantum read-only memory (QROM) \cite{babbush18} is one approach to storing and accessing data in quantum systems, and consists of an addressable $2^{n \times m}$-qubit state-generation circuit with $n$ qubits for address values that select one or more of the $N=2^n$ data words to be generated, and $m$ qubits for each word.
QROM is applicable to algorithms that do not require writing to memory, and it can provide quantum speedup by leveraging superposition to process multiple words of data simultaneously \cite{schuld18,ventura00,phalak22}.

\begin{figure}[ht]
  \centering
  \includegraphics[width=0.5\linewidth]{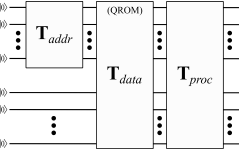}
  \caption{\label{fig:QROM_in_circuit}
  A generalized quantum algorithm with a QROM subcircuit. $\mathbf{T}_{addr}$ selects the addresses whose data is to be processed, $\mathbf{T}_{data}$ is the QROM state-generation circuit, and $\mathbf{T}_{proc}$ is the quantum algorithm that acts upon the loaded data.}
  \Description[QROM Subcircuit]{A series of horizontal lines representing qubits are intersected by rectangles, indicating the three parts of a quantum algorithm using a QROM subcircuit: the first rectangle marks the address-selection gates, the second rectangle marks the QROM itself, and the third rectangle marks the rest of the quantum procedure that will act upon the selected data.}
\end{figure}

Figure \ref{fig:QROM_in_circuit} consists of three segments and illustrates how a QROM subcircuit fits into a quantum algorithm.
$\mathbf{T}_{addr}$ selects the addresses whose data is to be processed, $\mathbf{T}_{data}$ is the QROM state-generation circuit, and $\mathbf{T}_{proc}$ is the quantum algorithm that acts upon the loaded data.
It is worth noting that the QROM portion, $\mathbf{T}_{data}$, and the data-processing algorithm portion, $\mathbf{T}_{proc}$, can be repeated for multiple rounds of data access.

There are several ways to synthesize QROM subcircuits, because while classical bits have only a single mechanism of storing data, qubits allow for more than one way of encoding information.
This is more clear in light of the multiple mathematical models for parameterizing qubits.
A qubit, $\ket{\Psi}$, can be modeled as 
\begin{math}
    \ket{\Psi}=\alpha\ket{0}+\beta\ket{1}
\end{math}, 
where the parameters $\{\alpha, \beta\}\in\mathbb{C}$ are probability amplitudes that specify the likelihood that $\ket{\Psi}$ will be measured in the associated state with respect to a given measurement basis.
A qubit may also be parameterized within the spherical coordinate system as
\begin{math}
    \ket{\Psi}=\cos\frac{\theta}{2}\ket{0}+e^{i\phi}\sin\frac{\theta}{2}\ket{1}
\end{math},
where $\theta$ represents longitudinal position on the Bloch sphere, and $\phi$ is the phase difference amongst the probability amplitudes, sometimes termed the phase of the qubit.
Several data encoding methods leverage these parameterized models differently \cite{schuld18,schuld21a,weigold20,ventura00}, and this section briefly introduces three types of encoding that can be used to create QROM.
Refs. \cite{sinha22} and \cite{schuld21} provide further details.

First, basis encoding restricts the probability amplitudes of a qubit to either $0$ or $1$, meaning that basis encoding restricts qubit usage to a one-to-one classical-bit-to-qubit mapping.
The mapping can be denoted in binary as 
\begin{math}
    (b_{m-1}b_{m-2}, \cdots,\allowbreak b_1b_0) \rightarrow \ket{b_{m-1}b_{m-2}, \cdots, b_1b_0} \text{, where } b_i \in \mathbb{Z}_2
\end{math}.
Second, angle encoding represents classical values using the parameterized qubit angles $\{\theta, \phi\}$.  
Thus, angle encoding requires all represented values to be interpretable as values within the range $[0,2\pi)$, but it allows for storing two $m$-bit values with a single qubit.
The mapping can be shown as 
\begin{math}
    x \in [0, 2\pi)\rightarrow cos(x_j)\ket{0}+e^{ix_{j+1}} sin(x_j)\ket{1}
\end{math}, 
where two classical data values are represented as $x_j$ and $x_{j+1}$.
Third and finally, improved angle encoding also represents classical values using the parameterized qubit angles $\{\theta, \phi\}$ \cite{sinha22}.
This encoding is `improved' with respect to the precision of values that may be represented, and it can be used in a method that is analogous to floating-point encodings for conventional electronic processing.
Unlike angle encoding, it stores only one $m$-bit value per qubit, because the angle $\theta$ represents the significand, $S$, of an $m$-bit value, and the phase angle $\phi$ represents an integer exponent, $E$.
Thus, a single $m$-bit value, $V$, is represented by a single qubit as $V=S\times2^{E}$, with the mapping
\begin{math}
    S, E \in [0, 2\pi)\rightarrow cos(S)\ket{0}+e^{iE}sin(S)\ket{1}
\end{math}.

Using these data encodings, a QROM can be mathematically represented as a set containing two $N$-dimensional vectors, denoted as $\{\vec{a}, \vec{x}\}$, where the $j^\text{th}$ components of each vector consist of the $n$-bit address field, $a_j$, and the $m$-bit data word, $x_j$. Consequently, every QROM represents $N=2^n\times m$ data entries, but how this information is stored depends upon the encoding choice, which is critical for best utilizing the qubits available on contemporary NISQ machines.
For example, the conventional approach of mapping a single classical bit to a single qubit is inefficient, and using alternative data encodings can improve the size of the QROM subcircuit in an application-specific manner \cite{sinha22}.

\subsection{Quantum Random Number Generation Subcircuits}\label{sub:random_number_generation}
\begin{figure}[ht]
  \centering
  \includegraphics[width=0.9\linewidth]{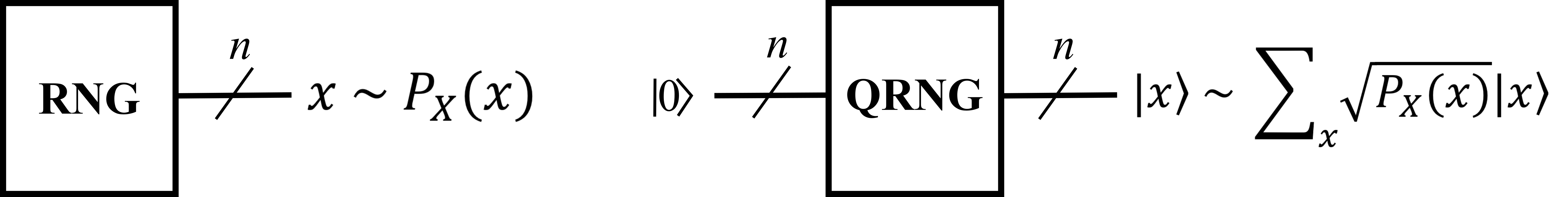}
  \caption{\label{fig:RNG_and_QRNG} The classical and quantum architecture of an RNG, where the former results in a probability distribution $P_X(x)$, and the latter results in a probability distribution of $\sum_X {\sqrt{P_X(x)}\ket{x}}$.}
  \Description[QRNG Subcircuit]{A cartoon contrasting an RNG (which is composed of a blackbox resulting in a probability distribution $P_X(x)$ and a QRNG, which is comprised of a blackbox resulting in a probability distribution of $\sum_X {\sqrt{P_X(x)}\ket{x}}$}
\end{figure}

Quantum random number generators are quantum subcircuits that generate statistical distributions, which can act as random number generators (RNGs).
Random number generation is critical in both classical and quantum applications.
First, RNGs are fundamental components of existing and emerging cryptographic methods, including post-quantum cryptography \cite{sinha23,barker07,turan18}.
Second, RNGs are necessary for modeling nonparametric distributions in simulations; for example, Monte Carlo methods require the rapid generation of RNGs.
Third, random number distributions are often used as components of quantum algorithms.
For example, they can be used as non-uniform priors in algorithms such as Grover's search, assigning different weights to regions of the search space based on the likelihood of finding a solution.
They are also important in QML algorithms, including Markov models, Boltzmann machines \cite{amin18}, and Bayesian networks \cite{low14}.
Figure \ref{fig:RNG_and_QRNG} illustrates the structure of a classical RNG and a QRNG.

Significant work has discussed and evaluated uniform, technology-independent QRNGs \cite{herrero17,tamura20,salehi22,jacak20}.
But QRNGs need not generate only uniform distributions; preparing state-generation circuits that represent PMFs for a multi-qubit quantum computer provides convenient, flexible, and high-quality QRNGs \cite{sinha23}.
Unfortunately, synthesizing a quantum subcircuit representing a desired PMF is typically a complex task, especially for arbitrary, nonparametric PMFs.
Consequently, this work describes a solution: synthesizing programmable QRNGs by leveraging amplitude encoding.
Amplitude encoding represents information in the probability amplitudes, $\alpha$ and $\beta$, of a qubit by applying rotation gates that adjust the probability that a set of qubits will be measured in either the $\ket{0}$ or $\ket{1}$ states \cite{grover02}.
This approach to QRNG generation requires only two user-specified inputs: the number of qubits to be used and a tabular representation of the desired PMF, which may be non-parametric.

Before moving on, it is worth contextualizing QRNGs in the broader classification of RNGs.
The most general RNG classification is that of pseudo-random number generators (PRNGs) and true random number generators (TRNGs) \cite{sinha23,barker07,turan18}.
While PRNGs are useful for generating short-term similarity to TRNGs, they are deterministic and thus reproducible given a piece of secret information \cite{sinha23,golub13,matsumoto98}.
Conversely, TRNGs are non-deterministic, and are thus appropriate when random values must be unpredictable in advance with a chance greater than fifty percent \cite{sinha23,barker07,turan18}. 
Natural quantum processes often provide the randomness for TRNGs, thus making it sensible to consider QRNG subcircuits as TRNG candidates \cite{barker07,turan18}.
However, QRNG subcircuits alone are often not TRNGs.
For example, measurement devices add deterministic biases to sampled output, making QRNGS weakly random sources (WRS).
Consequently, extractors---mathematical functions formulated alongside RNG distributions---are used to remove the deterministic components from weakly random sources.
This paper discusses solely QRNGs in their WRS form, meaning absent extractors, because non-uniform extraction techniques are a separate, complex topic \cite{ma13}.

\subsection{Quantum Oracle Subcircuits}\label{sub:quantum_oracles}
Oracles are subcircuits used in several notable quantum algorithms, from the Deustch-Jozsa algorithm to Shor's Prime Number Factorization algorithm, as well as in several important quantum subroutines such as quantum phase estimation \cite{deustch92,cleve98,bernstein97,grover96,simon97,shor94,harrow09,nielsen11}.
While oracles differ in their exact functionality, they are all `blackbox functions' that filter a quantum circuit's statespace.
Oracle functions are thus mathematical functions that either transform inputs in a user-specified way to `mark' them as solutions, or output those same inputs without change when they are not solutions.

Consequently, the process of specifying oracle subcircuits begins with specifying a mathematical function that satisfies three sets of constraints.
The first is dictated by the problem the quantum algorithm is designed to solve.
Considering a conceptually-straightforward example of Grover's Search algorithm, the function to be synthesized must adjust any input that satisfies a certain search requirement while leaving unchanged any inputs that do not.
Furthermore, because most quantum algorithms use qubits that are defined as states over a binary basis, most functions of interest must be represented as `switching functions'---functions whose input and output values are designated with solely zero's and one's---prior to formulating an oracle subcircuit for the function.

\begin{figure}[ht]
  \centering
  \includegraphics[width=0.75\linewidth]{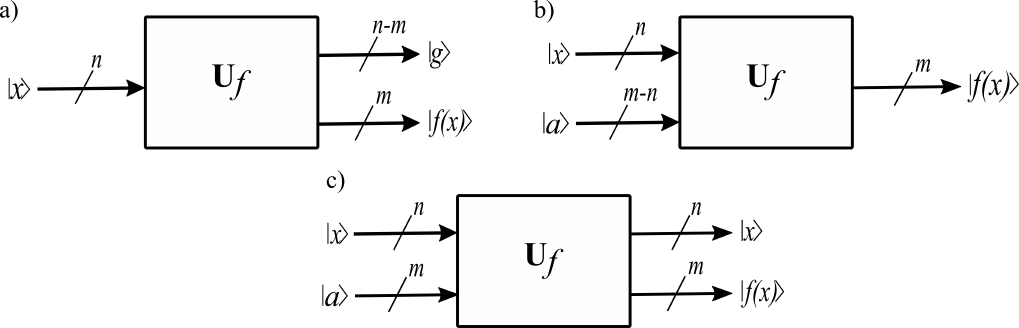}
  \caption{\label{fig:oracle_architecture}Three oracles; Subfigure a is an oracle where the input values are specified with more bits ($n$) than the output values ($m$), meaning only a portion of the input qubits are overwritten with output values. Subfigure b is the opposite situation, meaning the input values are specified with fewer bits than the output values, so only a portion of the qubits used to hold outputs must be initialized with the input values. Subfigure c is an oracle that preserves the domain values.}
  \Description[Oracle Subcircuit Architecture]{Cartoons illustrating the structure of three oracles: the first two represent oracles that have minimal number of qubits, so either one set of inputs is split into two sets of outputs (when there are more inputs than outputs, meaning some of the outputs are garbage) or two sets of inputs merge into one set of outputs (when there are more outputs than inputs, meaning some of the inputs are ancilla). The last subfigure has both two sets of inputs and outputs because this illustrates an oracle that perserves domain values.}
\end{figure}

The second set of constraints is imposed by the dictates of quantum mechanics, which require that oracle subcircuits represent reversible functions, which must be bijective.
This is true regardless of the fact that functions of interest often exist as non-bijective functions.
Third and finally are constraints imposed by the quantum circuit of which the oracle subcircuit will be a part.
Some quantum algorithms---when translated to quantum circuits---will make use of the domain values to an oracle after the oracle subcircuit's completion; in this case, the domain values must be on qubits that are not changed by the gates comprising the oracle.
Conversely, other algorithms do not make use of the oracle subcircuit's domain values, allowing the qubits on which those values are input to the oracle subcircuit to be repurposed as the outputs of the oracle.
Using the latter approach, we can design oracles with the provably minimal number of qubits, while the former approach requires adding `ancilla' and `garbage' qubits to the input and output sides of an oracle subcircuit.
Figure \ref{fig:oracle_architecture} illustrates these structures.

\begin{table}[htbp]
\caption{Summary of Subcircuit Types\label{table:subcircuit_summary}}
\begin{center}
\begin{tabular}{ccc}
\toprule
Subcircuit & Classical State \\
\midrule
Basis QROM & $\{\vec{a},\vec{x}\}, a_j \in \mathbb{Z}_2^n, x_j \in \mathbb{Z}_2^m$ \\
\hline
Angle QROM & $\{\vec{a},\vec{x}\}$, $a_j\in \mathbb{Z}_2^n, x_j \in [0, 2\pi)$ \\
\hline
Improved Angle QROM & $\{\vec{a},\vec{x}\}$, $a_j\in \mathbb{Z}_2^n, S_j \in [0, 2\pi), E_j \in [0, 2\pi)$ \\
\hline
QRNG & $x \sim P_x(X), x \in \mathbb{Z}_2^n$ \\
\hline
Oracle (Fewer Outputs)  & $f:\mathbb{Z}_2^n \rightarrow \mathbb{Z}_2^m, m<n$ \\
\hline
Oracle (Fewer Inputs) & $f:\mathbb{Z}_2^n \rightarrow \mathbb{Z}_2^m, n<m$ \\
\hline
Oracle (Equal Inputs \& Outputs) & $f:\mathbb{Z}_2^n \rightarrow \mathbb{Z}_2^m, n=m$ \\
\midrule
Subcircuit & Quantum State \\
\midrule
Basis QROM & $\ket{\psi_x} = \sum_{j=0}^{N-1}{\ket{a_j}\ket{x_j}}$\\
\hline
Angle QROM & $\sum_{j=0}^{(N-2)/2} \ket{a_j} (cos(x_{2j+1})\ket{0}+e^{ix_{2j}}sin(x_{2j+1})\ket{1}) $\\
\hline
Improved Angle QROM & $\sum_{j=0}^{N-1} \ket{a_j} (cos(S_j)\ket{0}+e^{i E_j}sin(S_j)\ket{1} )$\\
\hline
QRNG & $\sum_{j=0}^{N-1} \sqrt{P_x(X)}\ket{x}$\\
\hline
Oracle (Fewer Outputs)  & $\mathbf{U}_f \ket{x}\ket{a} \rightarrow \ket{f(x)}$\\
\hline
Oracle (Fewer Inputs) & $\mathbf{U}_f \ket{x} \rightarrow \ket{g}\ket{f(x)}$\\
\hline
Oracle (Equal Inputs \& Outputs) & $\mathbf{U}_f \ket{x}\ket{a} \rightarrow \ket{g}\ket{f(x)}$\\
\bottomrule
\end{tabular}
\end{center}
\end{table}

\subsection{Situating Our Toolkit with Previous Work}
Substantial effort has gone into designing tools for efficient interfacing  with quantum circuits, and this has led to packages for directly implementing circuits on hardware, for simulating quantum hardware, and for situation-specific applications, such as quantum machine learning  \cite{serrano22,demicheli22}.
A detailed description of this collective work is beyond the scope of this paper, so we leave that to the thorough analysis of articles such as Refs. \cite{serrano22} and \cite{demicheli22}.
However, it is important to clarify how \textit{MustangQ} fits into this landscape of tools, and specifically to note that the goal of \textit{MustangQ} is not to replace the many sophisticated quantum software design tools that already exist (especially those with thousands of users, such as Qiskit, PyTket, and Cirq), but rather to complement them.

Specifically, \textit{MustangQ} differs from existing software packages in three ways.
First, it starts with classical functions, whereas most tools require that the user prepare a first representation of a function as a quantum circuit---or an equivalent mathematical representation---before that circuit can be further optimized or mapped to hardware.
Second, \textit{MustangQ} is built to be customizable on a level that many packages are not; it is designed so that C/C++ codes with different synthesis methods, optimizations, or mapping techniques may be `plugged in' via a module specifically designed to make adding new commands to the toolkit straightforward.
Third and finally, \textit{MustangQ} aims to be a connecting hub for tools that quantum circuit developers may not know how to use; for example, as described in Section \ref{sub:connecting_with_other_tools}, \textit{MustangQ} allows the user to indirectly interface with Qiskit functionality for transpiling circuits for specific hardware.

We also note that an empirical comparison of \textit{MustangQ}'s circuits to those of other other packages is beyond the scope of this work for three reasons.
First, as mentioned above, many quantum circuit packages do not take functions as input, but rather start with preliminary quantum circuits that are then adjusted or optimized.
So, to compare \textit{MustangQ}'s circuits to those from alternative packages, we would need to manually prepare starting quantum circuit representations for the latter, which is both beyond the scope of this work and not feasible for several of the functions we consider in Section \ref{sec:MustangQ_subcircuits}.
Second, even for those tools that do have the option of functions-as-input, it would be a substantial data processing task to convert the benchmark functions we use to formats required by other tools.
Third, as described in Section \ref{sec:introduction}, this paper does not consider implementation on quantum hardware, meaning that metrics such as computation fidelity on a given device are not available for circuit comparison in this work.
Consequently, while comparing \textit{MustangQ}'s outputs to those of other packages would be valuable and interesting, it is a task separate from this initial presentation of our toolkit.

\section{\textit{MustangQ}: A Quantum Synthesis, Compilation, and Optimization Toolkit}\label{sec:MustangQ}
\subsection{Introduction to \textit{MustangQ}}\label{sub:MustangQ_intro}
Our synthesis, compilation, and optimization toolkit---\textit{MustangQ}---has been in development for several years, and is currently in its fourth formative version \cite{smith19,smith19a}.
Given the growing need to efficiently design optimized quantum circuits, this toolkit is designed to be as general as reasonably possible, targeting both application-specific quantum integrated circuits (QIC) and more general quantum design automation (QDA) tools via synthesis, optimization, verification, and visualization \cite{smith19,smith19a,sinha22,sinha23,henderson23}.
\textit{MustangQ} supports a custom circuit representation structure that can currently operate with fifteen types of quantum gate (including parameterized gates); classical input preprocessing; five core synthesis methods; and several optimization and postprocessing methods, including the ability to generate circuit specifications that are ready to run on IBM's suite of cloud-accessible hardware.
While this paper focuses on \textit{MustangQ}'s ability to synthesize quantum subcircuits, we use this section to both introduce the toolkit's overall functionality (all of which is leveraged in Section \ref{sec:MustangQ_subcircuits}), and also to describe the directions in which \textit{MustangQ} is growing.

\subsection{Preparation for Synthesis: Preprocessing Classical Specifications}\label{sub:pla_format}
Because \textit{MustangQ} is designed to automate conversion of classical function specifications to quantum circuit representations, it is important to have a sufficiently-flexible classical specification format.  
Consequently, \textit{MustangQ} is capable of synthesizing quantum circuit specifications from irreversible and incompletely-specified conventional switching function descriptions.
(Recall that a switching function is one whose input and output values are designated with solely zero's and one's.)
\textit{MustangQ} can utilize several classical input formats, because the input specification processing is handled separately from circuit synthesis and optimization; thus, the toolkit is designed to work with inputs including gate-based netlists (written in formats such as Verilog or RevLib's REAL \cite{revlib11}) or decision diagrams (such as BDDs \cite{bryant86,wille09}).
While the tool has the ability to work with all of these file types, the functionality described here uses the \texttt{.pla} file format, which is a tabular approach \cite{brayton84,rudell86} that we selected as the exemplary format for two reasons.
First, \texttt{.pla} files are relatively straightforward and standardized, providing a user-friendly---and parser-friendly---way to specify classical functions.
Thus, \texttt{.pla} files are consistent with ease-of-use, a substantial goal of automating conversion from classically-specified functions to quantum circuits.
Second, we have a sizeable set of benchmark function specifications in either \texttt{.pla} format or in other tabular forms that are readily converted to \texttt{.pla} files, and this includes collections maintained via RevLib \cite{wille08}.
Consequently, the \texttt{.pla} format is convenient for toolkit development and evaluation, but it is not the only input format with which \textit{MustangQ} can or should operate.

Each \texttt{.pla} file represents a function with discretized domain and range values that are comprised of bitstrings.
Every row of a \texttt{.pla} file consists of $n+m$ bits: the first $n$ bits are one unique variable valuation in the function domain, and the subsequent $m$ bits represent the corresponding function valuation.
While \texttt{.pla} files can theoretically specify any function, full function specification requires exponential space to hold a table with $O(2^n)$ rows and $O(n+m)$ columns.
Thus, the \texttt{.pla} format includes three features to reduce the size of function specifications \cite{brayton84}.
First, $n$-bit function variable values that differ in only a single bit value from other variable values can be represented with a single row; a dash (``-'') indicates that both $0$ and $1$ values are acceptable for the replaced bit.
Second, undefined function domain values are not explicitly specified.
And third, dashes (``-'') can replace function range values that may be either $0$ or $1$.
Because the first and third features may seem duplicative, it is worth emphasizing how they differ: when a ``-'' is in one of the $n$-bit strings that represent function variable valuations, it indicates that \textit{each} of $0$ and $1$ can replace the ``-'', meaning two separate function variable valuations---and corresponding range values---are represented for every ``-'' present.
When a ``-'' is in one of the $m$-bit strings that represent function range values, it indicates that \textit{either} $0$ or $1$ is an acceptable value for the specified bit; these dashes are sometimes termed ``don't cares,'' because the function `doesn't care' about that bit of the output \cite{revlib11}.
While it is ideal in terms of reduced storage to utilize all of these table-reducing features to their fullest, some of the circuit synthesis methods require preprocessing that adjusts the \texttt{.pla} file contents; such adjustments include assigning values to the don't cares to prepare one-to-one or onto functions that embed the original functions of interest that may be irreversible or incompletely-specified.
We include details on \textit{MustangQ}'s preprocessing capabilities in Appendix \ref{app:preprocessing_details}; here, it is simply worth emphasizing that some quantum circuit synthesis methods included in \textit{MustangQ} operate without any preprocessing, and others operate with varying amounts of preprocessing.
This leads to different method strengths, such as those observed in Section \ref{sec:MustangQ_subcircuits}.

\subsection{Quantum Circuit Synthesis Methods}
This section presents five approaches to quantum circuit synthesis.
Within \textit{MustangQ}, circuits are constructed using a custom-designed data structure that represents qubits and the gates that act on them as a collection of related linked lists.
This allows for efficient construction and optimization of circuits with hundreds of thousands of quantum gates.
Of course, such circuits are far too complex for today's NISQ machines, and likely any machines of the foreseeable future.
Consequently, Section \ref{sec:MustangQ_subcircuits} will focus on construction and optimization of much smaller circuits, but as quantum hardware grows, it seems likely that the world's most sophisticated machines will soon be able to consistently run circuits with thousands of gates \cite{arute19,castelvecchi23,computing23}.
Therefore, it is worth noting that our toolkit is designed to accommodate substantially larger circuits.
\textit{MustangQ}'s default output for each of these synthesis methods is a technology-independent QASM specification termed OpenQASM \cite{cross17,cross22,cross23}; however, as is discussed in Section \ref{sub:optimization_and_post_processing}, other output formats are available.

\subsubsection{Exclusive-Or Sum of Products Synthesis}\label{sub:esop_synthesis}
Exclusive-Or Sum of Products (ESOP) synthesis acts on a classical function specified as an exclusive-OR sum-of-products.
Specifically, an ESOP is a function with inputs and outputs specified as bitstrings whose individual bits are interpreted as being combined via logical conjunctive AND operators.
These bitstrings (termed products) are then summed with disjunctive exclusive-OR operators \cite{fazel07}.
Evaluating the resulting expression with a given input bitstring produces that input's corresponding output bitstring.
Although most functions specified as \texttt{.pla} files will not natively be in ESOP form, they can be converted via an automated process, whereupon the rows in the \texttt{.pla} table are the products of the ESOP \cite{mishchenko01,henderson23}.

\begin{figure}[ht]
  \centering
  \includegraphics[width=0.6\linewidth]{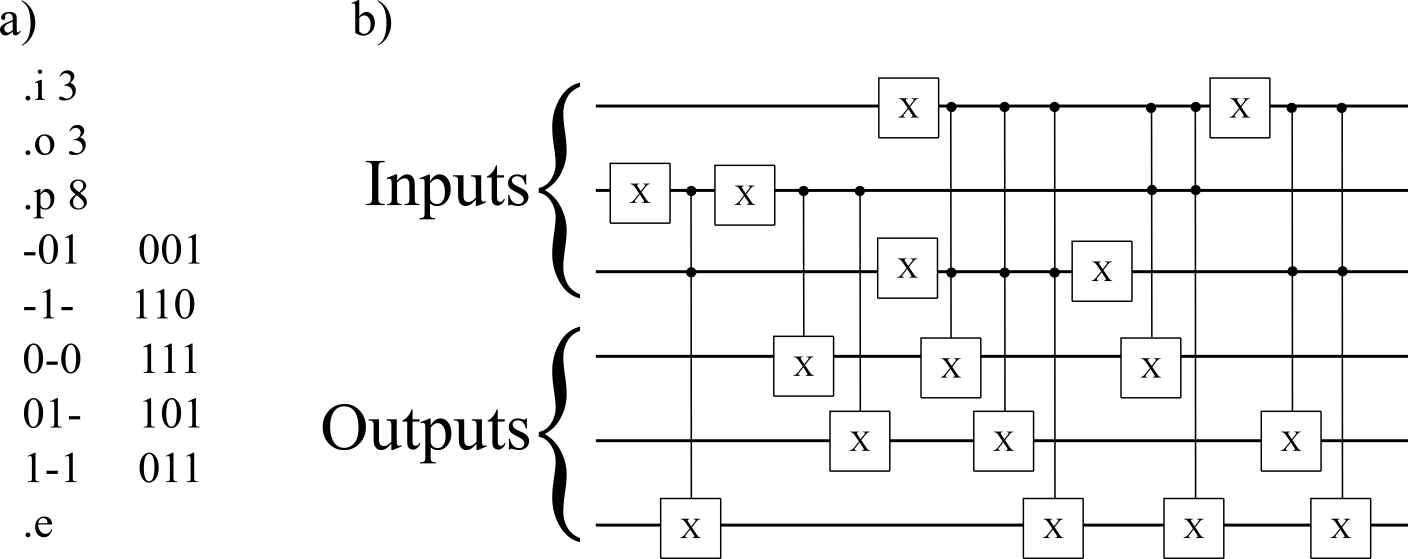}
  \caption{\label{fig:esop_example}Subfigure a represents a classical function with three inputs and three outputs in the \texttt{.pla} file format. Subfigure b contains an ESOP-synthesized circuit generated from that function.}
  \Description[Quantum Circuit from ESOP Synthesis]{A circuit comprised of six horizontal lines representing qubits with six Pauli-$\mathbf{X}$ gates, eight true (two-control) Toffoli gates, and two controlled-Pauli-$\mathbf{X}$ gates.}
\end{figure}

The ESOP synthesis method produces a quantum circuit from a \texttt{.pla} table in ESOP format by mapping each product to a Toffoli gate \cite{fazel07} in two steps.
First, qubits are added to the circuit: one qubit is added for each input variable, and one qubit is added for each output variable.
Qubits must be added for both the input and output variables, because circuits generated with ESOP synthesis do not overwrite the input values with output values.
Second, for each product in the table, and for each output variable value of $1$ therein, a Toffoli gate is added to the circuit.
The controls of the Toffoli gate are determined by the inputs of the current product: an input variable value of $1$ means a control is placed on the corresponding input qubit, while an input value of $0$ means a negative control is placed on the corresponding input qubit.
A negative control is simply a control with a Pauli-$\mathbf{X}$ gate on either side; the Pauli-$\mathbf{X}$ gates reverse the qubit's polarity before the control and restore it after the Toffoli gate has executed.
The Toffoli gate's target is placed on the output qubit corresponding to the output variable whose value is $1$.
Figure \ref{fig:esop_example} illustrates a quantum circuit created using ESOP synthesis.

\subsubsection{Transformation-Based Synthesis}\label{sub:TBS}
Transformation-Based Synthesis (TBS) is a family of quantum circuit synthesis methods described in works including Refs. \cite{miller03}, \cite{miller20}, and \cite{soeken16}.
All TBS methods produce reversible quantum circuits using the minimal number of qubits, and \textit{MustangQ} currently implements several variants.
For brevity, this paper discusses two: the foundational ``basic, unidirectional method,'' and an alternative approach that can generate less complex circuits by interpreting the function in a different way (see Section \ref{sub:TBS_RM}).

The basic unidirectional TBS method works for classical function specifications provided as truth-tables that are completely-specified, are reversible, and have input and output bitstrings of equal length.
(Using the notation of this paper, $n = m$.)
Consequently, most classical functions to be synthesized will require preprocessing, and---for any function that is represented in a truth-table format that can be converted to the \texttt{.pla} format described in Section \ref{sub:pla_format}---\textit{MustangQ} automates all processing via parsers and the methods described in Section \ref{sub:expand_assign_one_to_one} and \ref{sub:onto}.

\begin{figure}[ht]
  \centering
  \includegraphics[width=0.7\linewidth]{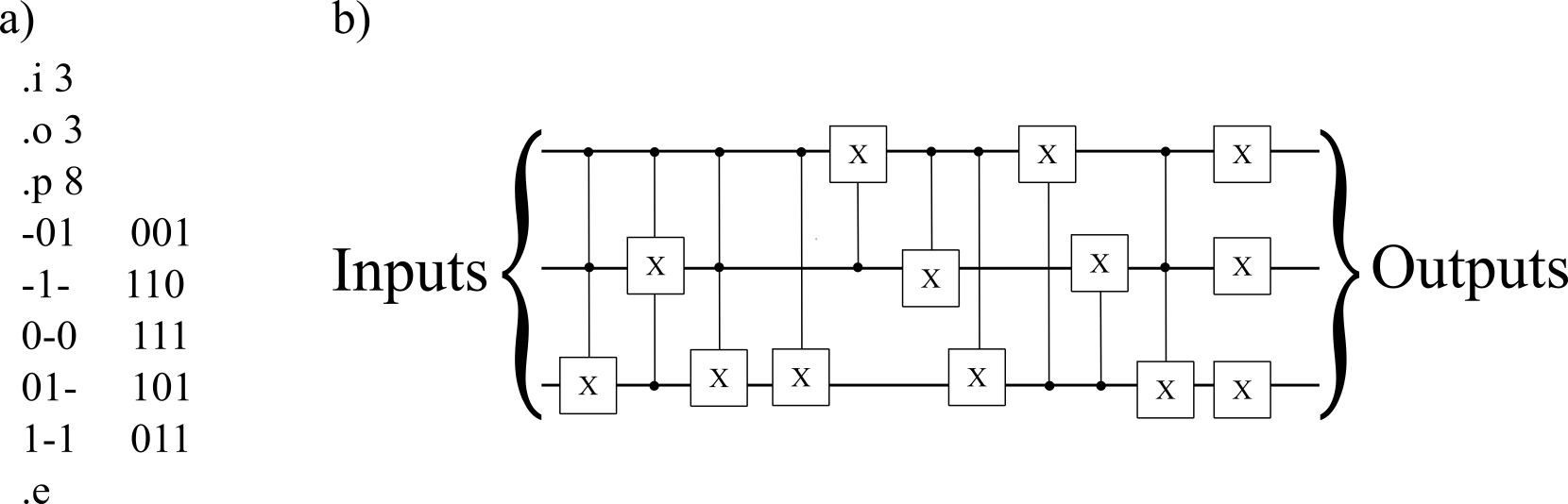}
  \caption{\label{fig:tbs_example}Subfigure a represents a classical function with three inputs and three outputs. Subfigure b represents a TBS circuit generated from that function.}
  \Description[Quantum Circuit from TBS]{A circuit comprised of three horizontal lines representing qubits with three Pauli-$\mathbf{X}$ gates, four true (two-control) Toffoli gates, and six controlled-Pauli-$\mathbf{X}$ gates.}
\end{figure}

The core of all TBS methods is to iteratively transform the input classical function into the identity function, and the reversed series of gates that achieves this is the output quantum circuit.
The unidirectional TBS method applies gates to the output side of a quantum circuit such that each output pattern in the classical specification is identical to its corresponding input pattern \cite{miller03}.
(Other approaches leverage the classical specification's reversibility to apply the method in both directions simultaneously, thereby choosing to add gates on one or the other end of the circuit to reduce the overall gate count \cite{miller03}.)
Specifically, TBS works through the given tabular specification in ascending order of the input patterns (\textit{i.e.,} $(0...00), (0...01), (0...10) ... (1...11)$), chooses the gates necessary to map each output to the corresponding input, and then applies those gates to the function being synthesized to determine an intermediate function that becomes the function to be synthesized in the next iteration.
The key to TBS is that gates are chosen such that they do not affect any entry earlier in the specification \cite{miller03}: the resulting quantum gate cascade is comprised of Pauli-$\mathbf{X}$ gates (\textit{i.e.,} NOT), controlled-$X$ gates (\textit{i.e.}, CNOT), Toffoli gates (\textit{i.e.,} controlled-controlled-$\mathbf{X}$ or C-CNOT), and multiple-control Toffoli gates (\textit{i.e.,} gates with three or more control inputs and a target Pauli-$\mathbf{X}$ operation) \cite{miller03}.
Figure \ref{fig:tbs_example} illustrates a quantum circuit created using TBS.

\subsubsection{Transformation-Based
Synthesis with the Reed-Muller Spectrum}\label{sub:TBS_RM}
Transformation-based synthesis with the Reed-Muller spectrum is a variant of TBS that was designed in hopes that it would make circuits less complex than those generated with TBS. 
To understand the method, we must first consider the Reed-Muller expansion, and we may begin by considering the expansion for a single-output Boolean function, $f(X)$.
Such a function may be expressed using the positive polarity Reed-Muller expansion or PPRM, given as
\begin{math}
f(X) = \bigoplus_{i=0}^{2^n-1} a_i P_i.
\end{math}
Here, $X = \{x_1, x_2, ..., x_n\}$, $a_i = \{0,1\}$, and $P_i$ is the function input $x_j$ where the $j$th positions of $i$ expressed in binary are $1$.
`Positive polarity' indicates that each of the $x_i$ terms appear non-negated throughout the function, but because the TBS-RM method uses solely this varation of the Reed-Muller spectrum, from hereon out, we use `Reed-Muller expansion' to refer to the positive polarity variant.
Given a tabular representation of a binary function with only a single output, the coefficients of the RM expansion---which are collectively termed the RM spectrum---can be computed using the following expression:
\begin{math}
A = M^n F\text{, where } M^0 = \begin{bmatrix} 1 \end{bmatrix}, M^k = \begin{bmatrix} M^{k-1} & 0 \\ M^{k-1} & M^{k-1} \end{bmatrix}\text{, and } A = \begin{bmatrix} a_0 \\ a_1 \\ \vdots \\ a_{2^n-1}\end{bmatrix}.
\end{math}

We may extend this computation to Boolean functions with multiple outputs, which allows for recursively computing the Reed-Muller spectrum values as described in Ref. \cite{steinbach22}.
The Reed-Muller spectrum has a number of important properties, and the most relevant here is that computing the $a_i$ for a given input row in a tabular function specification requires information from \textit{only} the preceding rows.

Define $R_{i,j}$ as the $i$th component of the Reed-Muller spectra for a function with multiple outputs, meaning that this is the $i$th component of the Reed-Muller spectra for $f_j(X)$.
The TBS-RM method then works as follows: after starting with an empty circuit, we apply a Pauli-$\mathbf{X}$ gate for every value in $R_{0k}$ that is $1$.
Then, for every other function input---indexed between $1$ and $2^n-1$---we take a different action, depending upon whether or not the index is a power of $2$.
If the index is a power of $2$, then consider $R_{ik}$; if $R_{ik}=0$, we make it $1$ by applying a controlled-Pauli-$\mathbf{X}$ gate, with the control on the qubit with index $s=\max{(j|R_{ij}=1, j>k)}$ and the target on $x_k$.
We then apply a second set of controlled-Pauli-$\mathbf{X}$ gates, for every $R_{ij}=1$ with $j\neq k$.
Note that, here, `applying' a gate means that we add the gate to the output side of the circuit, \textit{and} then update $R$ by transforming $R$ to the functional domain, applying the gate to obtain the updated $R$, and then transforming back to the Reed-Muller spectrum domain.

If the index is not a power of $2$, there are more steps.
First, we compute $s=\max{(j|R_{ij}=1)}$ such that the output with position $s$ is $0$ in the binary expression of the index for the current row ($i$).
Then, we apply controlled-Pauli-$\mathbf{X}$ gates with controls on each $s$ and targets on $j$, where $R_{ij}=1$ and $j\neq i$.
Next, we apply multiple-controlled Toffoli gates, with controls on $x_j$ such that the $j$th position of the binary expression of $i$ is $1$, and with targets on $x_j$.
Finally, for any gate that affected an input/output value that preceded the current value at index $i$, re-apply those controlled-Pauli-$\mathbf{X}$ gates to undo that effect.

Upon reversing the order of the resulting gate sequence, we have a quantum subcircuit that preserves input values and uses the minimal number of qubits, while also ideally using less expensive gates than the basic TBS variant.
TBS-RM can make circuits simpler, because it \textit{does} consider input/output pairs that have preceded the current pair whose Reed-Muller values are under consideration.
While the TBS-RM method adds gates later to ensure this `backward-looking' consideration does not affect the outcome of the circuit, the ability to so generally results in choosing gates that have fewer controls, making the circuit less complex.

TBS considers the function specification in a fixed order, and we have found that it is possible to organize the computation so that only one row of the specification is required for each iteration.
This has potential advantage for functions with a large number of inputs and outputs as the complete specification need not be stored.
We have found the same feature is possible for the TBS-RM
method: the Reed-Muller coefficients can be computed one at a time as needed using only the corresponding function entry.
This means that the complete spectrum does not have to be computed and stored and, most importantly, it is not necessary to jump between the function specification and the spectrum in the manner noted above.

\subsubsection{Introduction to Encoded Synthesis Methods}\label{sec:intro_to_encoding}
Before introducing the next set of synthesis methods, it is worth a few clarificatory notes about their structure.
Basis, angle, improved angle, and amplitude encoding synthesis are all methods that create quantum circuits that store certain data in different portions of a parameterized qubit.
Basis encoding stores the information in the qubit basis states, angle and improved angle in the $\theta$ and $\phi$ parameters, and amplitude in the probability amplitudes associated with the basis states.
In this form, only a single piece of data can be stored at any given time, as adding gates to store more data overwrites or changes the data previously stored.
Consequently, encoding methods are often used in concert with controlled gates that associate data values (whether stored via basis state, angle, or probability amplitude) with bitstring indexes that are set using the gate controls.
As will be illustrated in Section \ref{sec:MustangQ_subcircuits}, \textit{MustangQ} generates such circuits because they have useful applications, and the following sections thus describe synthesizing controlled, data-encoding circuits that associate index values with each piece of data to be stored.
Table \ref{table:data_encoding_summary} summarizes these data encoding types.

\begin{table}[htbp]
\caption{Summary of Data Encodings\label{table:data_encoding_summary}}
\begin{center}
\begin{tabular}{ccc}
\toprule
Encoding Type & Classical Data & Quantum State \\
\midrule
Basis Encoding & $x \in \mathbb{Z}_2^m$, $x=(b_{m-1}, b_{m-2}, \cdots,b_0)$, $b_j \in \mathbb{Z}_2$ & $\ket{x} = \ket{b_{m-1} b_{m-2} \cdots b_0}$ \\
\hline
Angle Encoding &  $x \in [0, 2\pi)$ & $\otimes_{j=0}^{N} cos(x_{j})\ket{0}+sin(x_{j})\ket{1} $\\
\hline
Amplitude encoding & $ x \in \mathbb{R}^{2^{n}}, \sum_{i=1}^{2^{n}}\left|x_{i}\right|^{2}=1$ & $\left|\psi_{x}\right\rangle=\sum_{i=1}^{2^{n}} x_{i}|i\rangle$ \\
\bottomrule
\end{tabular}
\end{center}
\end{table}

\subsubsection{Basis Encoding Synthesis}\label{sub:basis_encoding}
Basis encoding is the simplest approach, as it maps one classical bit to one qubit by setting each qubit to either the $\ket{0}$ or $\ket{1}$ state.
Despite its simplicity, basis encoding does have utility, including fewer preprocessing requirements and its use in several quantum algorithms, including the Quantum Fourier Transform (QFT).
The approach's simplicity, however, has the cost of requiring many qubits.
A basis encoding that uses controlled gates to associate indices with data can be synthesized using an approach very similar to Section \ref{sub:esop_synthesis}, albeit without the minimization.

\subsubsection{Angle and Improved Angle Synthesis}\label{sub:angle_encoding}
As mentioned in Section \ref{sub:quantum_memory}, angle encoding stores data in the $\theta$ and $\phi$ parameters of a qubit.
Variants of angle encoding use these angles in slightly different ways; for example, ``dense encoding'' stores two data values with a single qubit, while improved angle encoding stores only one value per angle to allow for further precision \cite{sinha22}.
But all variants require that the data fall within an interval of $2\pi$ such as $[0,2\pi)$ or $[-\pi,\pi)$, which often requires data preprocessing including normalization, as described in Section \ref{sub:normalization}.
The normalization used to achieve a desired range and precision can vary substantially, and is the primary difference between angle encoding and what we term improved angle encoding.
However, the process of storing appropriately-specified classical values within a quantum circuit does not change based on the normalization method, and therefore, this section applies to both angle- and improved-angle-encoded circuits.

\begin{figure}[ht]
  \centering
  \includegraphics[width=\linewidth]{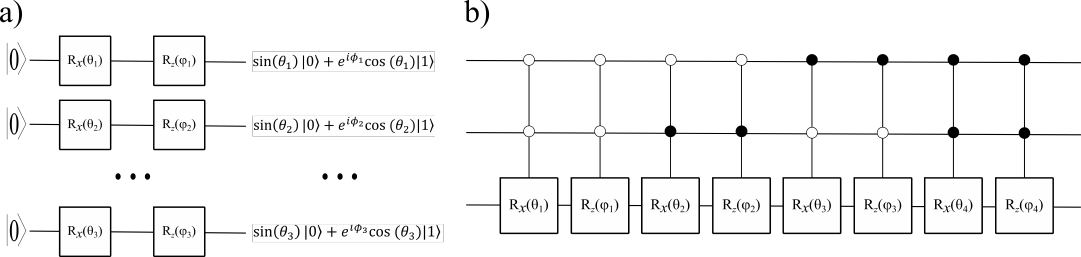}
  \caption{\label{fig:angle_encoding}Subfigure a illustrates the gates used to represent classical data in a quantum circuit via the method of angle encoding. Subfigure b illustrates how that data can be associated with index values by controlling the rotation gates. Specifically, the white controls are `negative' controls that indicate a Pauli-$\mathbf{X}$ gate on either side, and the example circuit here illustrates four data values, each of which is associated with one of the indices $00$, $01$, $10$, or $11$.}
  \Description[Quantum Circuits Illustrating Angle Encoding]{Two subfigures illustrating angle encoding; the first illustrates just the rotation gates storing data values in $\theta$ and $\phi$, while the second adds controls to those rotation gates to illustrate the structure of an angle-encoded circuit that associates each data value with an index.}
\end{figure}

The process is quite straightforward; it uses two types of rotation gates, $\mathbf{R}_x$ and $\mathbf{R}_z$, and the $\mathbf{R}_x$ gates specify the values stored within the probability amplitudes, while the $\mathbf{R}_z$ gates specify the phase angle.
Subfigure a of Figure \ref{fig:angle_encoding} illustrates this, while subfigure b illustrates controlling the rotation gates to create an index/value association.
Improved angle encoding has the same structure, except that it requires one data value to be stored using both $\theta$ (mantissa) and $\phi$ (exponent), instead of allowing for either one or two data values to be stored using these angles.

The primary benefit of angle encoding is that it requires far fewer qubits than basis encoding, as up to two data values, each consisting of $m$ bits, may be stored in a single qubit.  (As opposed to basis encoding, which requires $m$ qubits for $m$ bits.)
Furthermore, at the expense of only one $m$-bit value stored, improved angle encoding avoids underflow in situations where the difference between the largest and smallest values to be encoded is quite large.

\subsubsection{Amplitude Encoding Synthesis}\label{sub:amplitude_encoding}
Amplitude encoding stores information in the probability amplitudes of qubits, and it differs from basis and angle encoding in three important ways.
First, data values are always associated---at least implicitly---with indices.
This is a consequence of the fact that a probability must always be associated with an event.
Second, unlike basis and angle encoding, the data must be encoded `circuit-wide' and not qubit-by-qubit.
This will be further explained in the synthesis example below, and it again is a consequence of the fact that amplitude encoding uses \textit{probabilities}, which must satisfy certain circuit-wide constraints.
Third and finally, amplitude-encoded circuits do not have the same `uncontrolled' and `controlled' variants as basis- and angle-encoded circuits do; amplitude-encoded circuits must always utilize controlled gates.

\begin{figure}[ht]
  \centering
  \includegraphics[width=0.75\linewidth]{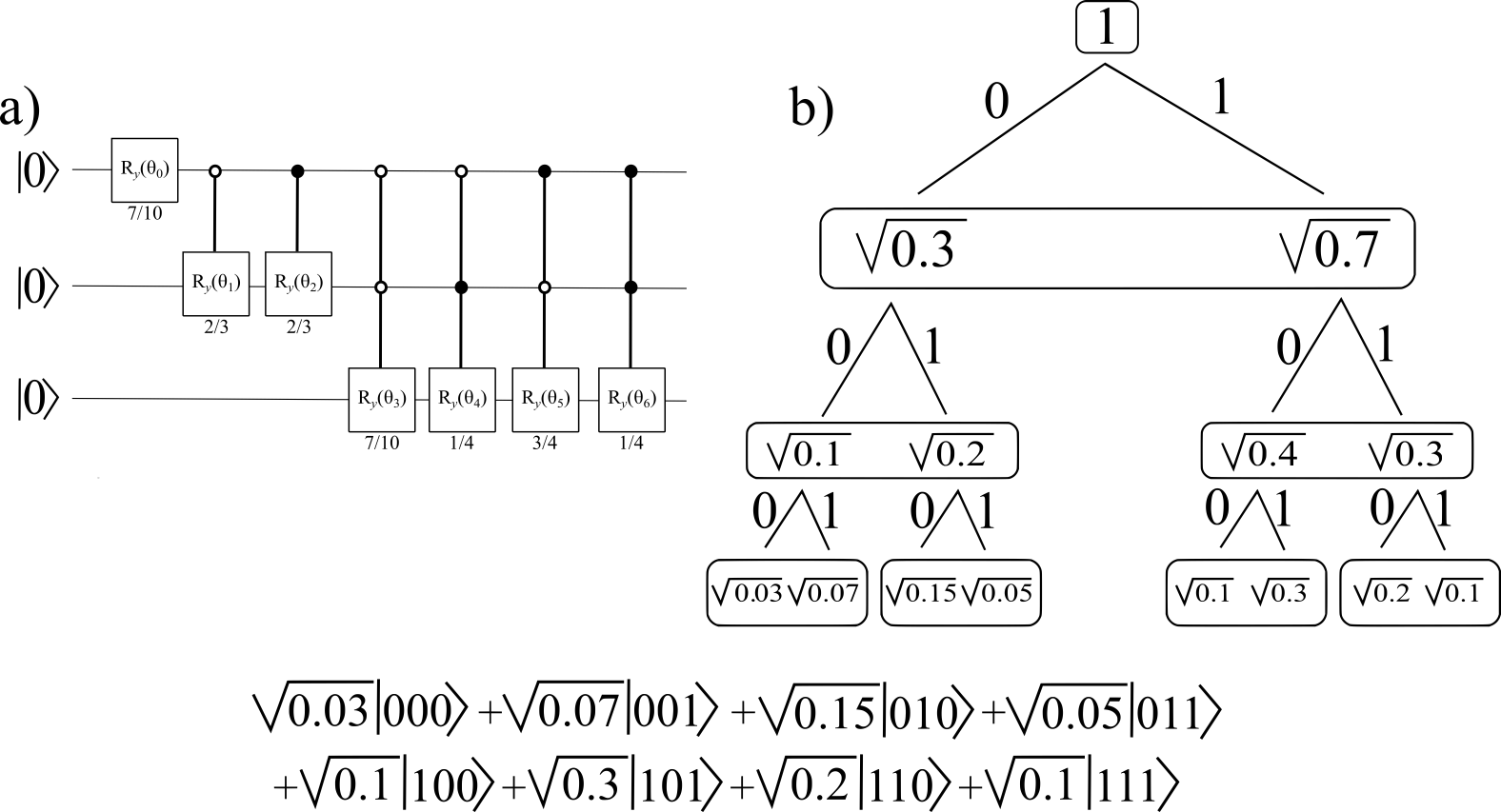}
  \caption{\label{fig:amplitude_encoding}An amplitude-encoded circuit and a representation of the information it holds. Subfigure a depicts the circuit itself, with controlled rotation gates, acting with rotations given by the $\theta_i$ values. Subfigure b is a binary-tree representation of the information in subfigure a: the weights along the edges form a bitstring `index' associated with the data value at each node. The expression beneath both subfigures lists each index/value pair in Dirac notation.}
  \Description[An Amplitude-Encoded Quantum Circuit.]{An amplitude-encoded circuit with three qubits, one single rotation gate, two single-controlled rotation gates, and four multi-controlled rotation gates alongside a tree diagram illustrating the probability distribution's relation to the generated circuit.}
\end{figure}

Amplitude encoding applies multiple-controlled rotations about the $y$-axis that tune qubits to have a certain probability of being in either the $\ket{0}$ or $\ket{1}$ state \cite{grover02}.
This synthesis procedure is easiest to explain with an example, so consider Figure \ref{fig:amplitude_encoding}, in which subfigure a illustrates an amplitude-encoded circuit, and subfigure b illustrates a binary tree containing the information represented by the circuit.
Each layer in the tree has unit-normalized node values, which contain the square roots of the probabilities of measuring the bitstring specified by the edges leading to the node.
For example, measuring the first qubit will give $0$ with $30\%$ probability and $1$ with $70\%$ probability, while measuring all three qubits will give $000$ with $3\%$ probability and $111$ with $10\%$ probability.
Consequently, Figure \ref{fig:amplitude_encoding} can be understood as encoding the data $0.03$ associated with `index' $000$ and the data $0.1$ associated with `index' $111$.

To determine the angles for constructing an amplitude-encoded circuit, begin with a specification of the index and data values, which can be represented in a tabular form such as that specified in Section \ref{sub:pla_format}.
Then, by iterating over the data values in a manner analogous to traversing the tree of subfigure b, we determine the rotations to be applied, as follows:
\begin{math}
  \theta_i = \arccos \sqrt{f(i)},
\end{math}
where each $\theta _i$ corresponds to creating a subtree of the ultimate bitstring index/probability distribution.
The function $f$ is given for a bitstring indexed by $i$ as
\begin{math}
  f(i)=\frac{\sum_{x_L^i}^{\frac{x_R^i-x_L^i}{2}} P(x) }{\sum_{x_L^i}^{x_R^i} P(x)}
\end{math},
where $x_L^i$ are the bitstrings associated with left-most node values of the subtree, $x_R^i$ are the bitstrings associated with right-most node values of the subtree, and the $P(x)$ is the probability amplitude associated with the bitstring $x$.
Each $f(i)$ thus obtains the sum of the probabilities on the left side over the range $[x_L^i,\frac{x_R^i-x_L^i}{2}]$ and divides it by the total probability in the full range of bitstrings $[x_L^i,x_R^i]$.

\subsection{Optimization and Post-processing}\label{sub:optimization_and_post_processing}
This section describes circuit operations that are designed to be applied after synthesis, and which---at least in most cases---may be applied in any order.
For example, irrelevant Pauli-$\mathbf{X}$ gates may be removed either before or after multi-controlled rotation gates are optimized with symmetric optimization, and circuit verification could happen---and probably ought to happen---after each of these steps.
As has been emphasized in other sections, refining quantum circuits to use fewer qubits and gates is of critical import for near-term hardware and likely for all hardware of the immediate future \cite{kim23}, so we have hopes to extensively expand the post-processing abilities of \textit{MustangQ} in the future.
In the meantime, this section describes the optimizations that are currently provided for application on the custom circuit representations of \textit{MustangQ}, the post-processing tools for verifying and evaluating synthesized circuits generated in \textit{MustangQ}, and the features of other quantum software packages that can be leveraged via connection to \textit{MustangQ}.

\subsubsection{Gate Reduction and Decomposition}\label{sub:reduction_and_decomp}
One of the most common sets of optimizations removes needless gates, including consecutive Pauli-$\mathbf{X}$ gates in groups of twos, which, as with classical NOT gates, amount to no change.
\textit{MustangQ}'s double Pauli-$\mathbf{X}$ removal has been shown to effectively reduce circuit gate count, as illustrated in Section \ref{sec:MustangQ_subcircuits}.

\begin{figure}[ht]
  \centering
  \includegraphics[width=\linewidth]{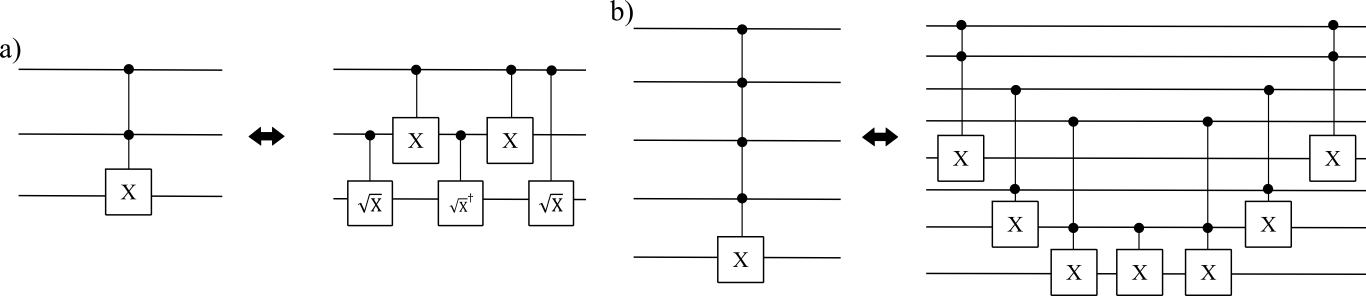}
  \caption{\label{fig:toffoli_decomposition}Decomposing Toffoli gates into constituent gates. Subfigure a illustrates decomposing a true (two-controlled) Toffoli gate into two controlled-Pauli-$\mathbf{X}$ gates, as well as two controlled-square-root-of-NOT gates, and a controlled-square-root-of-NOT-dagger gate.  Subfigure b illustrates decomposing a multi-controlled or generalized Toffoli gate into six true Toffoli gates and a controlled-Pauli-$\mathbf{X}$ gate.}
  \Description[Decomposing Toffoli Gates.]{Two circuit optimizations: the first shows breaking a true (two-controlled) Toffoli gate into two controlled-Pauli-$\mathbf{X}$ gates, as well as two controlled square-root-of-NOT gates, and a controlled square-root-of-NOT-dagger gate.  The second shows breaking a multi-controlled or generalized Toffoli gate into six true Toffoli gates and a controlled-Pauli-$\mathbf{X}$ gate.}
\end{figure}

Additionally, given the variety of limited gate sets currently available on different quantum hardware, \textit{MustangQ} can automate the translation of synthesized circuits from one gate set to another, including via decomposition of complex gates.
As Toffoli gates are not implemented natively on many quantum computers, 
\textit{MustangQ} implements two types of Toffoli gate decomposition: decomposing a generalized Toffoli gate (with more than two controls) into a sequence of `true' Toffoli gates (with two controls each) as in subfigure a of Figure \ref{fig:toffoli_decomposition} \cite{barenco95}, and decomposing a true Toffoli gate into the sequence of five controlled gates shown in subfigure b of Figure \ref{fig:toffoli_decomposition} \cite{barenco95}.
While these decompositions increase qubit count and gate count, they allow \textit{MustangQ} to remove complex controlled gates that cannot be natively run on quantum hardware.

To produce circuits compatible with a specific gate set, \textit{MustangQ} can also leverage the capabilities of the Qiskit library to transform one circuit into an equivalent circuit with a user-specified set of quantum gates.
Specifically, a circuit loaded into \textit{MustangQ}'s custom data structure can be converted into a Qiskit circuit specification.
Using methods of the Qiskit library, this circuit can be converted to another Qiskit circuit that uses a specific set of gates.
Finally, \textit{MustangQ} can read a Qiskit circuit specification into its custom data structure.
And so, with the steps of this process occurring under-the-hood, \textit{MustangQ} can produce circuits with alternative gate sets.

\subsubsection{Graycode Optimization}\label{sub:graycode_opt}
Graycode optimization is applicable to circuits with multi-controlled rotation gates.
Multi-controlled gates are generally amongst the most expensive \cite{nielsen11}, and Graycode optimization improves circuit performance by reducing the number of controls on consecutive multi-controlled rotation gates \cite{sinha22,daskin13}.

Graycode optimization leverages the fact that---in many cases---the order in which multi-controlled rotation gates are applied will be irrelevant.
Because the target rotations are controlled, they will definitionally be applied correctly regardless of whether the circuit `checks' the set of controls that applies first, last, or anywhere in between.
Consequently, we can apply fewer controls by re-ordering the multi-controlled gates to leverage information about shared control values.
Specifically, the gates are rearranged using a Gray code, which---in the classical realm---is a method for ordering bitstrings such that adjacent values differ by only a single bit \cite{black20}.

\begin{figure}[ht]
  \centering
  \includegraphics[width=0.8\linewidth]{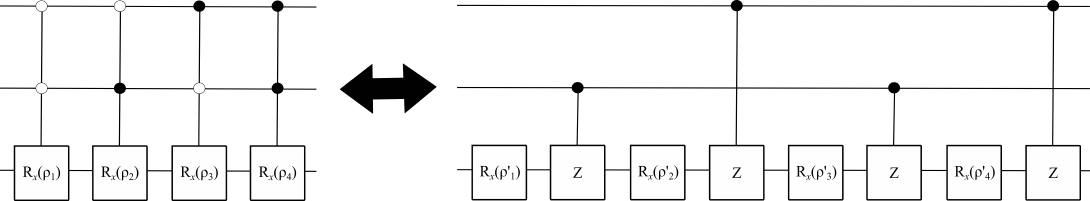}
  \caption{\label{fig:graycode_optimization}Reduction of multi-controlled gates using Graycode optimization.}
  \Description[Illustrating Graycode Optimization.]{Illustrating conversion of a circuit with four multi-controlled rotation gates into a circuit with eight gates overall, but with only single-qubit and single-controlled gates.}
\end{figure}

In a quantum circuit context, the Graycode optimization converts $2^n$ $n$-controlled rotations about the $x$-axis to $2^n$ total gates comprised of controlled-$\mathbf{Z}$ gates and single-controlled rotations about the $x$-axis.
This procedure computes the updated rotation values using the results of Refs. \cite{daskin13}, and details of both that computation and the optimization procedure are further discussed in Ref. \cite{sinha22}.
Figure \ref{fig:graycode_optimization} illustrates how Graycode optimization transforms four multi-controlled rotation gates to only single-qubit and single-controlled gates.

\subsubsection{Symmetric Circuit Optimization}\label{sub:symmetric_opt}
\begin{figure}[ht]
  \centering
  \includegraphics[width=0.75\linewidth]{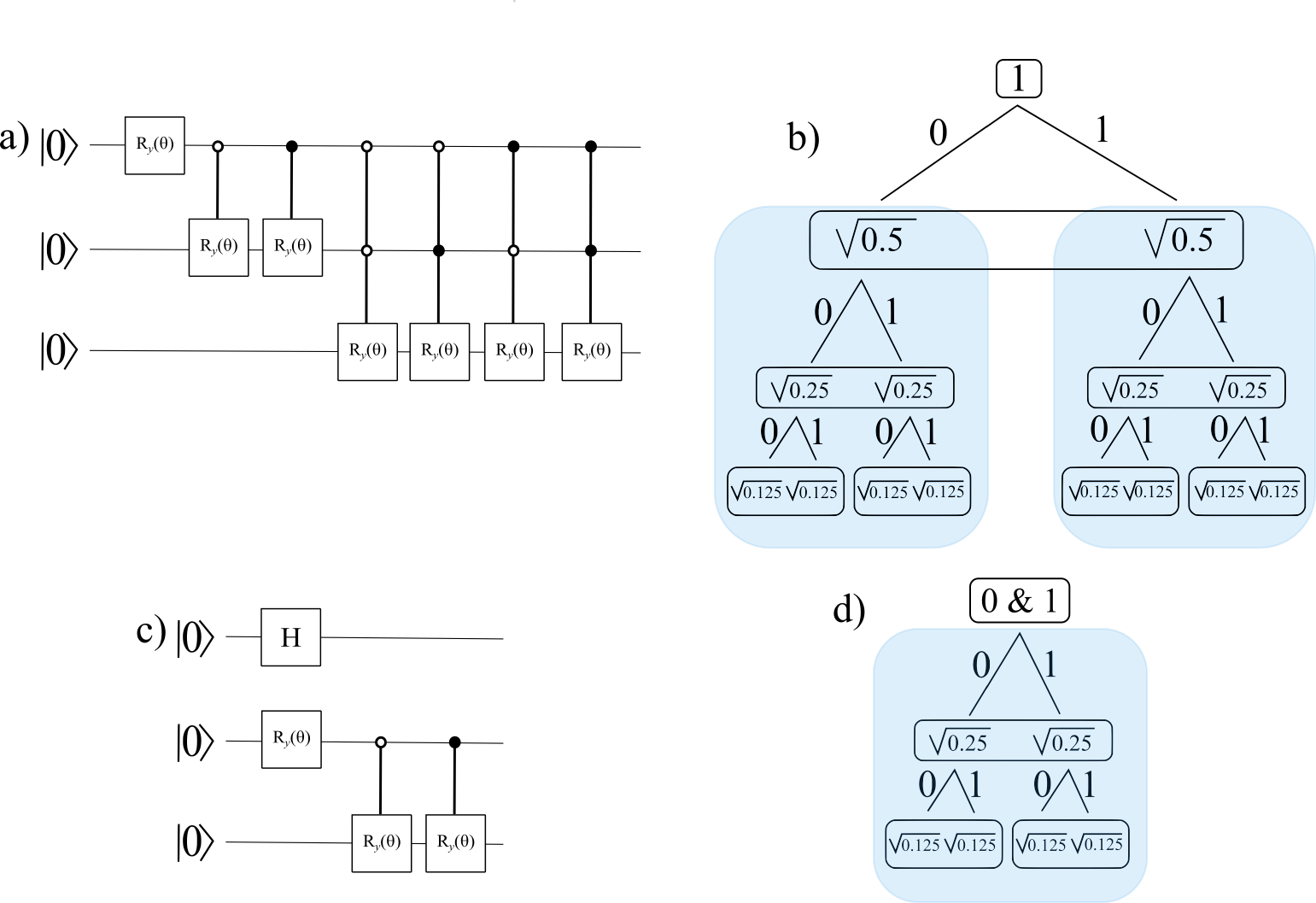}
  \caption{\label{fig:symmetric_duplicate}Leveraging the symmetric structure of a classical function to reduce gates in a quantum circuit. Subfigure a is the circuit associated with a duplicative function illustrated in subfigure b. Applying a Hadamard gate---and the resulting superposition---halves the number of gates in the circuit (subfigure c), such that the tree representing it contains a superposition of values leading to the duplicated portion (subfigure d).}
  \Description[An optimization that leverages symmetric structure of a classical function to reduce circuit gate count]{Circuits illustrating how a circuit containing four multi-controlled gates that is based upon a function with duplicative structure can be reduced to a circuit containing just two single-controlled gates.}
\end{figure}

\begin{figure}[ht]
  \centering
  \includegraphics[width=0.75\linewidth]{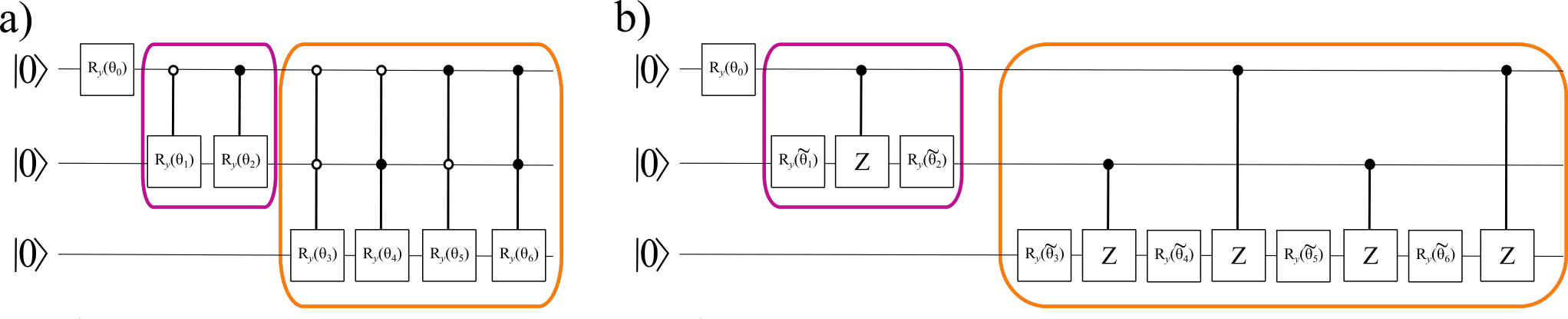}
  \caption{\label{fig:symmetric_duplicate_amplitude}Leveraging the symmetric structure of a classical function to reduce gates in a quantum circuit. Subfigure b is an optimized version of Subfigure a that uses the same principles illustrated in Figure \ref{fig:symmetric_duplicate}, but for an amplitude-encoded circuit, and not an angle-encoded one.}
  \Description[An optimization that leverages symmetric structure of a classical function to reduce circuit gate count]{Circuits illustrating how a circuit containing four multi-controlled gates that is based upon a function with duplicative structure can be reduced to a circuit containing just four single-controlled gates. This differs from the previous figure because it starts with an amplitude-encoded (and not an angle-encoded) circuit.}
\end{figure}

Symmetric circuit optimization leverages structural features of a classical function to reduce gate count in the resulting quantum circuit.
Specifically, for a classical function that has a symmetrical structure and that will be synthesized using either angle or amplitude encoding, this optimization can reduce the number of gates by half for each symmetry present.
It works by applying a Hadamard gate to one qubit, thereby `copying' the circuit that's constructed on qubits below, and this is most easily understood with an example.
Consider subfigure a of Figure \ref{fig:symmetric_duplicate}, which depicts an angle-encoded circuit for the data represented as a binary tree in subfigure b.
Because the highlighted portions of subfigure b's tree are identical, the circuit can be reduced to that in subfigure c of Figure \ref{fig:symmetric_duplicate}.
This is possible because the Hadamard gate puts the first qubit into a superposition of the $\ket{0}$ and $\ket{1}$ states, such that the remainder of the circuit (representing bits beyond the first in the bitstring) is represented with both values of the first bit: either $0$ or $1$.
Consequently, the gate count is reduced nearly by half, and we see how multiple symmetries in larger trees could have a repeated gate-reduction effect.
Subfigures a and b of Figure \ref{fig:symmetric_duplicate_amplitude} illustrate the same optimization for an amplitude-encoded circuit.

\begin{figure}[ht]
  \centering
  \includegraphics[width=0.75\linewidth]{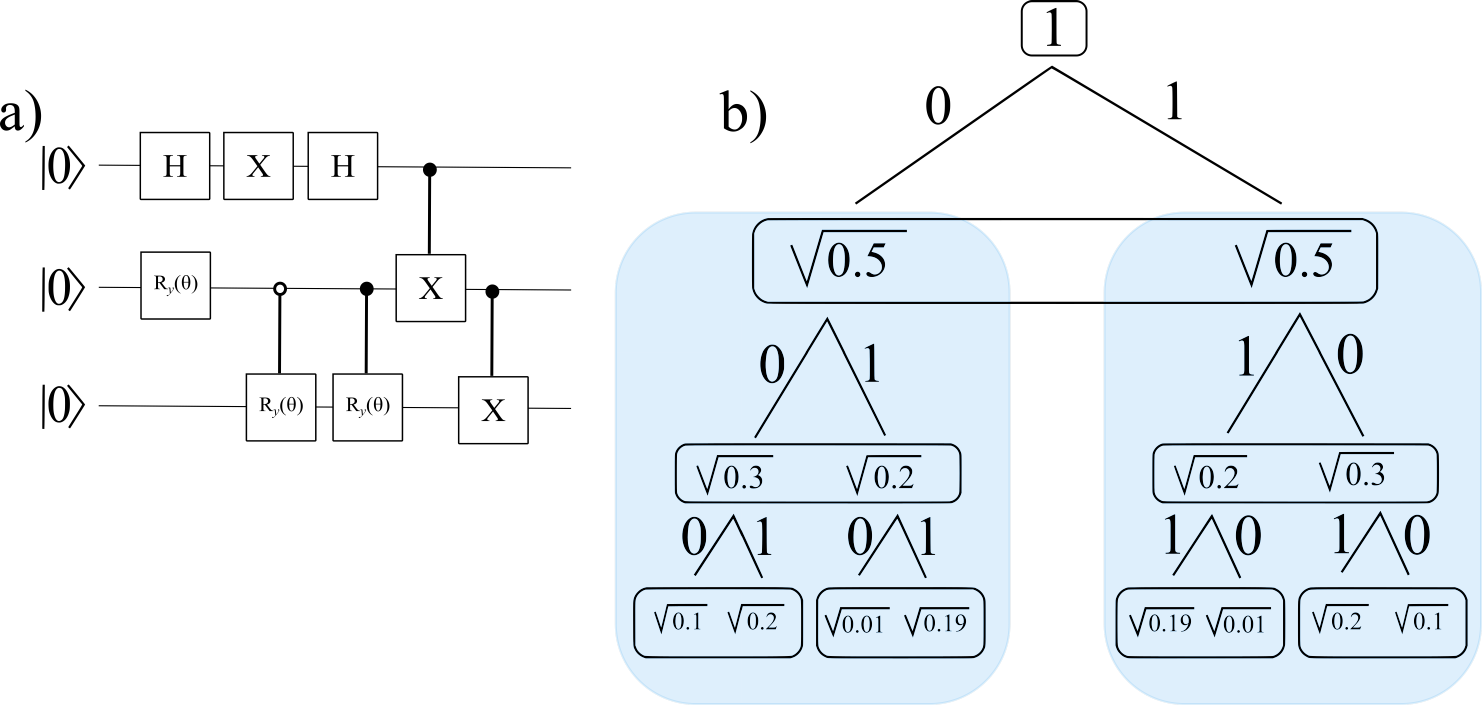}
  \caption{\label{fig:symmetric_mirror}Leveraging the symmetric structure of a classical function to reduce the number of multiple-controlled gates and the number of gates on one qubit in a quantum circuit. Subfigure a is the circuit associated with a mirrored function illustrated in subfigure b. Applying additional gates repeats the remainder of the function (as in Figure \ref{fig:symmetric_duplicate}) and mirrors it to produce a function illustrated in subfigure b.}
  \Description[An optimization that leverages symmetric structure of a classical function to reduce the number of multi-controlled gates and the number of gates on a qubit]{Circuits illustrating multi-controlled-rotation gate reduction for functions built from a tree diagram that has a structure illustrated in the associated tree diagram.}
\end{figure}

Just as we can leverage symmetric duplication of portions of classical functions, we can leverage symmetric mirroring of the same.
For functions that have mirrored portions (illustrated by the highlighted subtrees in subfigure b of Figure \ref{fig:symmetric_mirror}), we can apply a Hadamard gate and controlled-Pauli-$\mathbf{X}$ gate that will repeat the rest of the function, and a Pauli-$\mathbf{X}$ gate that will mirror it.
While this does not reduce the gate count,  it does reduce both the number of multiple-controlled gates and the number of gates that act on a single qubit, both of which contribute to simpler quantum circuits that are more likely to run to completion without decohering.

\subsubsection{Connecting with Other Tools}\label{sub:connecting_with_other_tools}
While \textit{MustangQ} may be used as a stand-alone toolkit for technology-independent circuit synthesis and optimization, it also aims to provide technology-dependent capabilities.
Hardware-specific operations are as tedious to perform manually as technology-independent ones, so automating these tasks is important, 
Consequently, previous versions of \textit{MustangQ} included custom mapping algorithms for target quantum hardware, including that of Rigetti and IBM \cite{smith19a}.
However, over the last half-decade, several available tools have been developed for sophisticated---and always-up-to-date---hardware-specific mapping and optimization \cite{qiskit23b,cirq23,sivarajah20}.
As it would be a mistake for \textit{MustangQ} to re-invent what these tools already provide, \textit{MustangQ} no longer supports its own technology-specific mappings and optimizations.
Instead, it provides users the ability to automate these tasks, without having to know how to use other quantum programming interfaces.

Currently, \textit{MustangQ} includes commands that allow users to apply Qiskit operations to circuits that were designed and optimized using \textit{MustangQ}.
Specifically, users may send circuits to Qiskit's visualizer, may decompose to Qiskit-provided gatesets for IBM hardware, or may prepare either OpenQASM or Qiskit specifications that are ready to run on IBM hardware and simulators.
Expanding the tools with which \textit{MustangQ} can interact is one of our top priorities, and specifically, we would very much like for users to be able to take advantage of Quantinuum's sophisticated Tket software \cite{sivarajah20}.
Tket interfaces with over twenty different technology-specific interfaces, and would allow users to easily prepare their \textit{MustangQ}-synthesized and optimized circuits for a panoply of quantum computer implementations without having to leave the \textit{MustangQ} environment.

\subsubsection{Quantum Circuit Verification and Evaluation}\label{sub:verification_evaluation}
As in the classical realm, quantum circuits must be verified to ensure that optimizations or other modifications do not change the overall circuit effect.
Thus, previous work has addressed the issue of quantum circuit verification \cite{miller06,niemann15,li22,tao22}, and \textit{MustangQ} aims to provide similar automated verification of the circuits in its framework.
Currently, it can verify some circuits for mathematical  correctness, it can compute statistics to assess the effect of sampling error, and it can compute statistics to approximate effects of hardware error.
The latter evaluation metrics are important because contemporary hardware is severely impaired by the effects of noise resulting from interference between qubits, interference between qubits and the environment, and qubit decoherence \cite{nielsen11,preskill18,kim23}.

First, formal verification.
\textit{MustangQ} currently has the ability to use Quantum Multiple-valued Decision Diagrams (QMDDs) \cite{miller06,niemann15} to verify circuits that have been optimized by removing or replacing gates, as long as the circuits being compared have the same number of qubits and use an appropriate gate set.
As described in Ref. \cite{smith19}, QMDDs provide an efficient data structure for manipulating the large matrices that mathematically represent quantum circuits, and they can thus also be used to formally verify modifications to such circuits.

Another part of formal verification is ensuring that generated circuits adhere to the desired specification.
Up to this point, this process for \textit{MustangQ} has consisted primarily of significant manual validation of reasonably-sized function specifications.
Additionally, for some synthesis methods, we have been able to verify circuits using software such as Verilog simulation scripts \cite{ieee08} or Qiskit simulation; the former can be used for circuits with only basis-encoded information---such as those created using ESOP synthesis of Section \ref{sub:esop_synthesis}---while the latter can be used for circuits with continuous gates, such as parameterized rotation gates.

It is worth noting that circuit verification is one component for which we would like to further significantly expand \textit{MustangQ}'s capabilities; for example, we would like to implement formal verification of circuits with differing numbers of qubits---that result from decompositions such as those in Section \ref{sub:reduction_and_decomp}---as well as circuits containing more gate types---such as general, parameterized rotation gates.

Second, evaluation.
We begin by clarifying the difference between sampling and hardware error: while the former is a feature of the probabilistic nature of quantum computing that would exist on even a perfect physical realization, the latter is a consequence of the current, imperfect devices that must work extraordinarily hard to avoid decoherence, and is thus theoretically removable \cite{nielsen11}.
Because \textit{MustangQ} can prepare circuits that are executable on specific hardware or hardware simulators, it can also collect statistics assessing the similarity between the results of these technology-dependent circuits and those of the classical circuit specifications from which the quantum circuits were prepared.
Specifically, for classical circuit specifications that are appropriately normalized---and which may thus be interpreted as a probability distribution---\textit{MustangQ} can compute distribution similarity metrics that assess how close a quantum circuit's distribution is to the desired distribution, even in the presence of sampling and hardware error \cite{mcdonald09}.\footnote{While these statistics might also be instructive for evaluating circuits that represent classical function specifications without normalized outputs, we do not further discuss such evaluation in this paper.}
Currently, \textit{MustangQ} can compute the Kullback-Leibler divergence, the Jensen-Shannon divergence, G-statistics, and associated G-statistic p-values.
The remainder of this section briefly introduces these concepts.

The Kullback–Leibler divergence (KL divergence) and Jensen-Shannon divergence (JS divergence) both measure the amount of difference between two distributions; the closer to $0$ the divergence, the more similar the distributions.
Specifically, KL divergence measures how different a discrete probability distribution, $Q$, is from a theoretical distribution, $P$, and is given by 
\begin{math}\label{eq:kl_divergence}
    \operatorname{KL}(P \| Q)= \sum_i P(i) \ln\left(\frac{P(i)}{Q(i)}\right)
\end{math} \cite{kullback51}.
The Jensen-Shannon divergence (JS divergence) is based upon the KL divergence, but is symmetric, in that it takes the average of the KL divergence when considering the distance between the empirical and theoretical distributions, and vice versa:
\begin{math}\label{eq:js_divergence}
    \operatorname{JSD}(P \| Q)=\frac{1}{2} KL(P \| M)+\frac{1}{2} KL(Q \| M) \text{ where } M=\frac{1}{2}(P+Q)    
\end{math} \cite{lin91}.
JS divergence thus remedies two issues with the otherwise-intuitive KL divergence: KL divergence is not a true distance metric, because it is not symmetric, and KL divergence can be undefined in cases where the empirical distribution does not overlap with the theoretical distribution.

A Goodness-of-fit test (or G-test) is another measure of the amount of similarity between two distributions, and is proportional to KL divergence:
\begin{equation}\label{eq:G_stat}
\begin{split}
G &= 2 \sum_i O_i \ln\left(\frac{O_i}{E_i}\right) \\
&= 2 \sum_i n P(i) \ln\left(\frac{n P(i)}{n Q(i)}\right) \\
&= 2n \sum_i P(i) \ln\left(\frac{P(i)}{Q(i)}\right) \\
&= 2n \times KL(P \| Q)
\end{split}
\end{equation} \cite{mcdonald09}.
Consequently, we can interpret KL divergence as a G-test statistic for assessing the difference between the distribution that a quantum circuit in fact represents and the ideal distribution that it is designed to represent \cite{mcdonald09}.
Furthermore, G-statistics have the benefit of being normalized (meaning always between $0$ and $1$, inclusive), which can provide a more intuitive sense of how likely it is that two distributions are the same.
This is especially true when a G-statistic is considered alongside a G-test p-value, which can be used since the G-distribution is approximately equal to the Chi-square distribution \cite{mcdonald09}.
Consequently, G-statistics are related to more-familiar `p-values': a G-statistic of zero is equivalent to a p-value of one, which strongly indicates that one \textit{cannot} say that one distribution is different from a second.
Another way to use these metrics is to note that both the G-statistic and its associated p-value provide a `similarity metric' between two distributions: when the G-statistic is close to zero and its corresponding p-value is close to one, it is more likely that two distributions are the same, and vice versa \cite{mcdonald09}.
In this work, we will term the p-value associated with a G-statistic a `similarity metric,' because it provides an intuitive measure of how similar two distributions are: if the similarity metric is zero, the distributions are likely to be completely different, and if the similarity metric is one, they are likely to be completely the same.
Anything in between indicates that the two distributions overlap to some degree.

\section{Automating Subcircuit Synthesis with \textit{MustangQ}}\label{sec:MustangQ_subcircuits}
This section illustrates \textit{MustangQ}'s ability to automate the creation of quantum subcircuits of three specific types: QROMs, QRNGs, and oracles.
While Section \ref{sec:MustangQ} provides a detailed description of our toolkit's raw capabilities, the goal of this section is to show those abilities in practice.
Specifically, we seek to highlight four aspects of \textit{MustangQ}: its ability to synthesize and optimize circuits far larger than could be created manually (e.g., $O(10^2)$ qubits and $O(10^3)$ gates) in a reasonable amount of time (e.g., $O(10)$ seconds); its ability to prepare circuits in a representation that is ready to run on existing quantum hardware or hardware simulators; the ability to compare between multiple synthesis and optimization approaches while using a single, unified computing toolkit; and the ability to facilitate novel research discoveries when raw synthesis is not the entire goal of a project.

Before going further, it is worth three clarificatory notes about our experimental results.
First, we chose to define a circuit complexity metric and to code it into \textit{MustangQ}'s statistics collection, because several non-standardized terms for measuring the size/difficulty of executing a quantum circuit exist.
This is a consequence of several factors, including differing considerations when considering technology-dependent versus technology-independent circuits, metrics that are heuristically-defined (such as quantum depth) \cite{qiskit23}, and terms that are oft-used but not to measure the same thing (such as `quantum cost' \cite{smith19,maslov05}).
Therefore, this section defines circuit complexity as the sum of the costs of each gate, where gate cost is the number of qubits on which a gate acts, including both control and target qubits.
While we make no claims as to the universal utility of this definition, it serves our purposes by reflecting both the number and size of gates in different \textit{MustangQ} subcircuits, regardless of the exact gate set represented in any given circuit.
Second, we use Qiskit's definition---and implementation---of quantum depth \cite{qiskit23}.
Third, we consider a sizeable set of benchmark function specifications, including from RevLib \cite{wille08}.
Such functions have `standardized' names that---although not semantically meaningful to those unfamiliar with them---are useful for identifying the functions that we used.
Consequently, we did not change the function names when reporting the circuit statistics, but in Appendices \ref{app:QROMs_data}, \ref{app:qrng_data}, and \ref{app:oracle_data}, we provide input and output counts for each function, to provide some quantitative information about each function.

\subsection{Quantum Read-Only Memories}\label{sub:MustangQ_QROMs}
\textit{MustangQ} can synthesize QROM subcircuits that use basis, angle, or improved angle encoding.
While a user inputs the same classical function specification to \textit{MustangQ} regardless of encoding type (as described in Section \ref{sub:pla_format}), the synthesis steps vary depending upon the data encoding applied.

First, the classical input (a function specified as a table of bitstrings) is preprocessed.
\textit{MustangQ} interprets the input as a memory image with $n$-bit address fields and $m$-bit data words, which is then expanded and assigned as described in Section \ref{sub:expand_assign_one_to_one}.
Furthermore, the classical function specification is made onto by assigning a data value of $0$ to any unrepresented addresses.
At this point, synthesis with basis encoding requires no further preprocessing.
However, synthesis with angle or improved angle encoding requires normalization: the $m$-bit memory words are interpreted as decimal values in either the range $[0, 1)$ or $[0,4)$, respectively, and are then normalized as described in Section \ref{sub:normalization}.
Following preprocessing, the second step is synthesis, which occurs using the methods described in Sections \ref{sub:basis_encoding} or \ref{sub:angle_encoding}.
Third and finally, the circuits can be optimized and post-processed using any of the methods described in Sections \ref{sub:reduction_and_decomp} and \ref{sub:graycode_opt} and \ref{sub:connecting_with_other_tools}.
The resulting circuit is a QROM subcircuit that can leverage address values in superposition to access corresponding superimposed memory words.

We synthesized circuits from twelve classical functions in a variety of ways to illustrate \textit{MustangQ}'s ability to synthesize QROM subcircuits, to compare the synthesis approaches for so doing (and illustrate how \textit{MustangQ} aids in such investigations), and to show \textit{MustangQ}'s ability to generate circuits with gate sets appropriate for contemporary hardware.
Specifically, we generated six circuits for each function: three with each of the encoding methods, and three with each of these methods plus optimization (double Pauli-$X$ gate removal for basis-encoded circuits and Graycode optimization for angle- and improved-angle-encoded circuits).
We further considered each of these circuits with both a `natural' gate set most applicable to each of the encoding methods, and also a uniform gate set that is native to IBM's quantum computers.
For basis encoding, the `natural' gate set consists of the Pauli-$\mathbf{X}$ gate and the two-controlled (\textit{i.e.,} `true') Toffoli gate, while for angle and improved angle encoding, it consists of multi-controlled rotation gates and single rotation gates.
The uniform gate set consists of single rotation gates and the controlled-$\mathbf{X}$ gate, and we were able to obtain the circuits in this format using the approach of Section \ref{sub:reduction_and_decomp}.
For all of the resulting circuits, we evaluated qubit count, gate count, circuit complexity, and quantum depth.
Appendix \ref{app:QROMs_data} reports the quantitative data; here we discuss the two most significant findings: the effect of optimization and the costs of basis encoding.

\begin{figure}[ht]
  \centering
  \includegraphics[width=\linewidth]{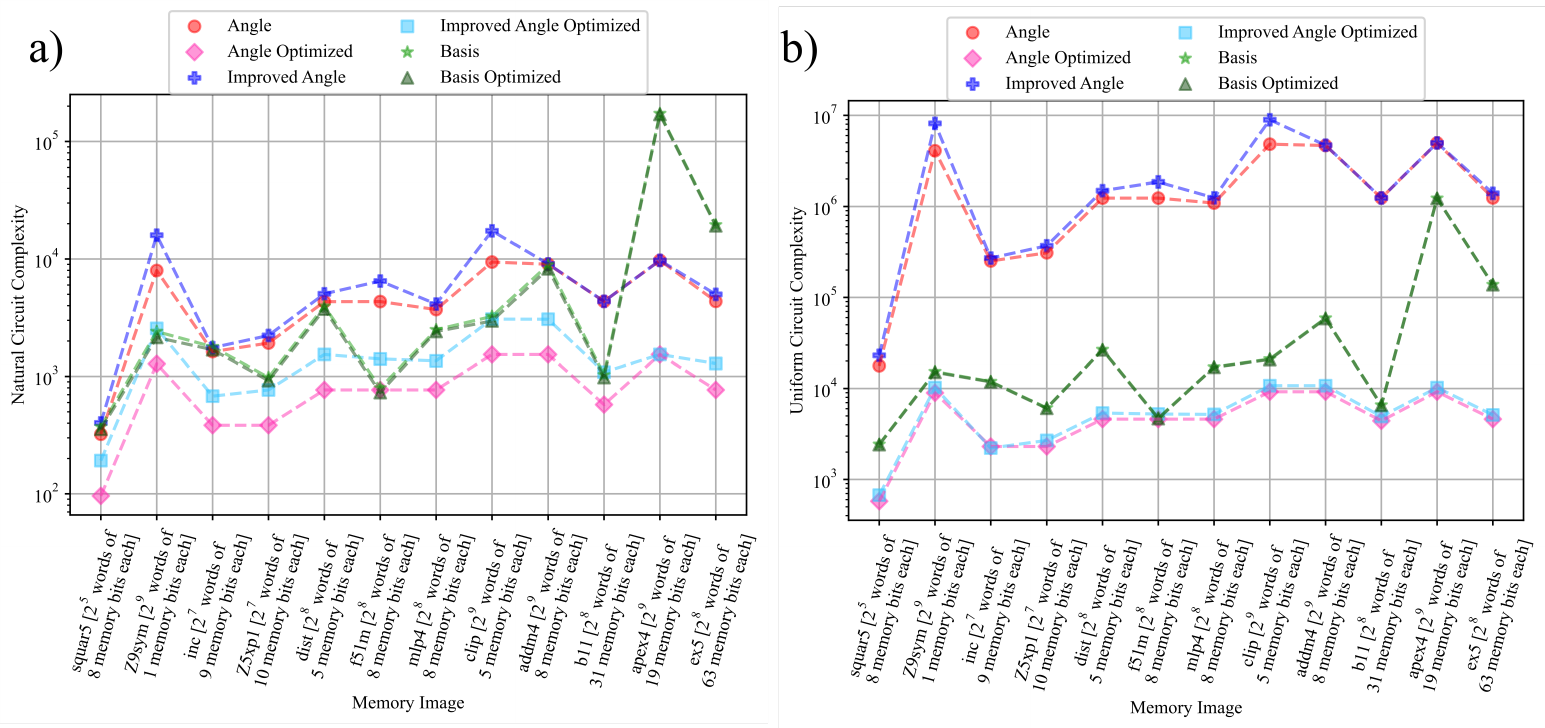}
  \caption{\label{fig:qrom_circuit_complexity}Circuit complexity for twelve benchmark functions synthesized using basis encoding, angle encoding, and improved angle encoding. Subfigure a illustrates results using the natural gate sets of each encoding, and subfigure b illustrates results using a uniform gate set.}
  \Description[QROM Subcircuit Circuit Complexity]{Two scatter plots illustrating the circuit complexity of twelve QROM subcircuits generated with three encoding methods each and with two different gate sets.}
\end{figure}

Unsurprisingly, optimization improves the synthesized circuits across all encoding types, and in both natural gate set and uniform gate set variants.
Of more interest is the fact that different encoding approaches benefit to varying degrees from the optimizations applied.
Subfigure a of Figure \ref{fig:qrom_circuit_complexity} illustrates that while natural gate set, basis-encoded circuits experience only a $6\%$ circuit complexity reduction when optimized, natural gate set angle- and improved-angle circuits experience $81\%$ and $66\%$ reductions, respectively.
The effect of optimization on uniform gate set encoded circuits is also noteworthy.
As illustrated in subfigure b of Figure \ref{fig:qrom_circuit_complexity}, converting circuits to a uniform gate set applicable for given quantum hardware generally increases the circuit complexity.
However, while the average complexity for angle- and improved-angle-encoded circuits increased by two orders of magnitude when converted from the natural to the uniform gate sets, the average complexity increased by only one order of magnitude with converted \textit{optimized} angle- and improved-angle-encoded circuits.

\begin{figure}[ht]
  \centering
  \includegraphics[width=0.6\linewidth]{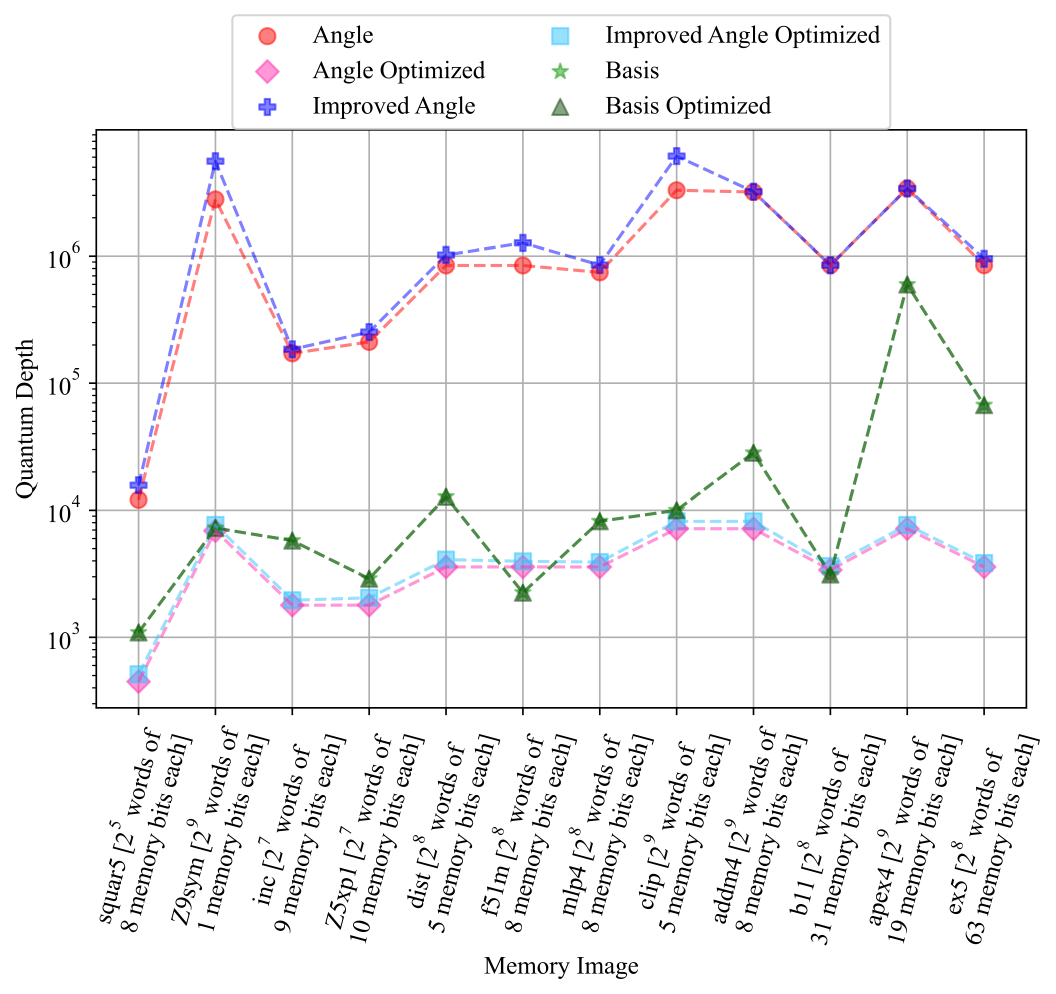}
  \caption{\label{fig:qrom_depth}Quantum depth for twelve benchmark functions synthesized using basis encoding, angle encoding, and improved angle encoding.}
  \Description[QROM Subcircuit Quantum Depth]{A scatter plot illustrating the circuit complexity of twelve QROM subcircuits generated with three encoding methods each.}
\end{figure}

As shown in Figure \ref{fig:qrom_depth}, the effect for quantum depth is similar, but even more pronounced; on average, optimization offers no depth reduction for basis-encoded circuits, while it reduces the average quantum depth of angle- and improved angle-encoded circuits by $99\%$ in both cases.
These results, of course, do not consider all possible optimizations, suggesting that identifying, implementing, and assessing the efficacy of additional optimizations is an important avenue of further research.

Comparing the encoding approaches illustrates that basis encoding---although arguably the most oft-applied---is generally the most expensive in terms of qubits, circuit complexity, and quantum depth.
At a minimum, basis encoding requires at least one qubit for each address and data value bit, and it can require several more when the generalized Toffoli gates are decomposed to two-controlled Toffoli gates, an optimiziation that is required for today's quantum hardware \cite{monz09,fedorov11}.
Consequently, on average, angle- and improved angle-encoded circuits have $61\%$ fewer qubits than basis-encoded circuits.
Similarly, on average, angle-encoded and improved-angle encoded circuits have circuit complexities that are $63\%$ lower and $29\%$ lower than those for basis-encoded circuits, with improved-angle encoding naturally having a lesser benefit given its requirement of more gates to offer better precision.
Finally, the same trend holds for quantum depth, where angle and improved angle encodings offer depths $44\%$ and $38\%$ less than basis encoding's average depth.

These findings about basis encoding are important, because they suggest avenues for future research.
Specifically, many quantum algorithms---including Grover's search algorithm and the Quantum Fourier Transform---utilize basis encoding.
This is likely because basis encoding is the most straightforward, but its relative expense suggests that it is important to learn whether and how quantum algorithms can leverage subcircuits with alternative data encodings.

\subsection{Quantum Random Number Generators}\label{sub:MustangQ_QRNGs}
\textit{MustangQ} can generate QRNG subcircuits with either parametric or non-parametric probability mass function (PMF) distributions.
Parametric distributions are those that can be defined with a simple set of parameters, such as the mean and the standard deviation, and non-parametric distributions are those for which the information cannot be so straightforwardly summarized \cite{corder14}.
Classical random number generation is generally limited to parametric distributions, as classical non-parametric distribution methods are often quite inefficient.
For example, the approach of rejection sampling requires generating substantial data that is discarded \cite{flury90}.
Consequently, \textit{MustangQ}'s ability to synthesize subcircuits for either parametric or non-parametric distributions is significant.

The first step in generating both parametric and non-parametric circuits is for the user to prepare a tabular specification (see Section \ref{sub:pla_format}) of a PMF for a desired distribution.
Both distribution types can be represented in this manner, since even non-parametric distributions can be explicitly represented as a set of `bins' and associated probability amplitude `heights.'
The number of bins is $2^N$ for a circuit that should use $N$ qubits.
For parametric distributions, Python's \texttt{SciPy} library can generate bin and corresponding height values for several common distributions using its \texttt{scipy.stats} functions \cite{virtanen20}.
For non-parametric functions, which are only defined by their user-specified bin and height relationships, the user prepares a table analogous to the one that \texttt{SciPy} prepares.

Having defined a tabular specification of the desired PMF, \textit{MustangQ} can address the remainder of the synthesis process.
It first preprocesses the PMF specification so that bin heights are interpreted as probabilities, which requires normalization such that the sum of the squares of the values is one, as described in Section \ref{sub:normalization}.
The normalized PMF specification is then used to synthesize a QRNG subcircuit as described in Section \ref{sub:amplitude_encoding}; specifically, the amplitude-encoding circuit structure stores the probability values of the PMF as probability amplitudes of quantum states \cite{sinha23}.
After synthesis, the circuit can be optimized and post-processed using the approaches of Section \ref{sub:reduction_and_decomp}, \ref{sub:graycode_opt}, \ref{sub:symmetric_opt}, and \ref{sub:connecting_with_other_tools}.
The result is an automatically-generated QRNG subcircuit that can be generated given only a distribution in the form of a PMF table and the number of qubits.

Unlike the QROM subcircuits of Section \ref{sub:MustangQ_QROMs}, which demonstrated \textit{MustangQ}'s scalability, here we demonstrate the ability to prepare subcircuits that have sufficiently few qubits to be run on contemporary hardware.
The QRNG subcircuits provide a good example, because even random number generators with only five qubits ($2^5=32$ discretized PMF bins) can be quite meaningful---and non-trivial to design---particularly if they are non-parametric distributions.
Therefore, we used \textit{MustangQ}'s ability to connect with Qiskit's backend optimizer and mapping tools to prepare circuit specifications that targeted the IBM suite of machines.
Specifically, we sought to assess the quality of \textit{MustangQ}-generated, optimized, and hardware-mapped QRNG subcircuits for five distributions (uniform, binomial, triangle, bimodal, and arbitrarily-specified), each of which was designed for a subcircuit with five qubits.
For each distribution, we used two sets of experiments to assess how well the QRNG subcircuit modelled the desired PMF and how well each withstood the effects of simulated hardware noise.
The first set of experiments used an ideal simulator that accounted for sampling error, but not hardware error, while our second set of experiments used a simulator designed to mimic the noise signature of a particular piece of existing quantum hardware \cite{ibmq23a}.\footnote{The choice of simulators is worth two further clarificatory notes. First, we chose to use simulators solely given the cost of accessing quantum hardware with sufficiently many qubits; having established preliminary results using simulators, we hope to extend these results to hardware in future work. Second, by `ideal simulator,' we do \textit{not} refer to what is sometimes called a `statevector' simulator that does not provide probabilistic output. Rather, we refer to a probabilistic simulator that does not include added inaccuracy to model hardware noise.}

To assess the circuits' quality, we had to ensure that inadequate sampling did not affect our results; thus, we computed a number of shots that would reduce sampling error.
While there are several metrics that can be used to compare circuit execution results to an ideal PMF, a Goodness-of-fit-test (or G-test) is one of the most easily interpretable, as described in Section \ref{sub:verification_evaluation}.
Because G-tests provide a normalized similarity metric for assessing how similar a theoretical distribution is to an empirically observed one \cite{sinha23,mcdonald09}, we considered only that metric here, but Ref. \cite{sinha23} provides further detail in how these quantities can complement each other.
We applied a G-test between each ideal PMF that should be represented and the results from executing a QRNG subcircuit on an ideal simulator; in each simulation, we made increasingly many shots until the G-statistic was less than $10^{-3}$, which indicates a p-value close to one for each distribution.
Out of an abundance of caution, we then increased that number of shots by $50\%$, and used it as the number of shots for our QRNG subcircuit assessment experiments.
Appendix \ref{app:qrng_data} reports all numerics for this section, including shot counts.
We found that---as might be expected---the more variation a distribution has, the more shots are required to reduce the effect of sampling noise.

\begin{figure}[ht]
  \centering
  \includegraphics[width=0.9\linewidth]{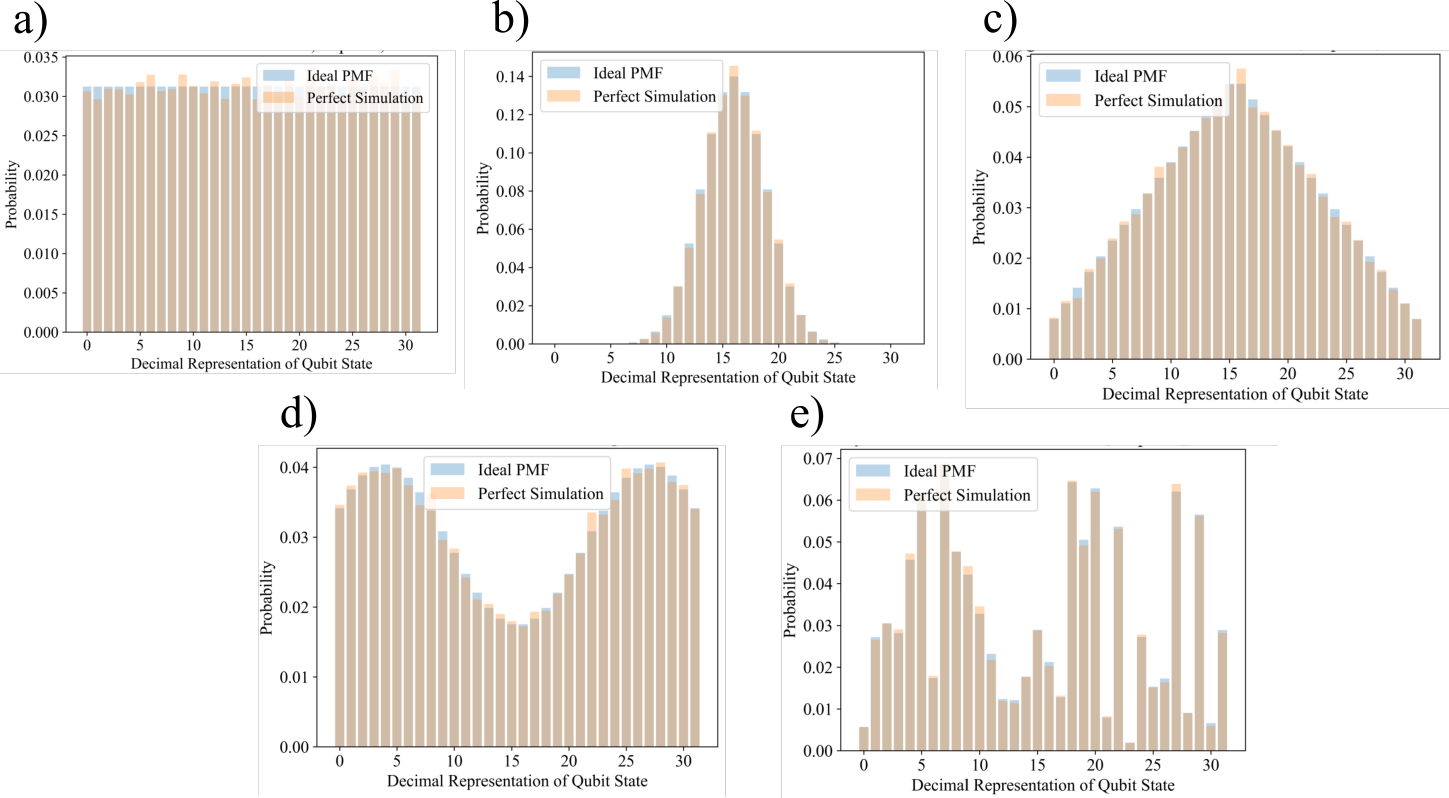}
  \caption{\label{fig:qrng_noiseless_simulation}Results of simulating five QRNG subcircuits using an ideal (\textit{i.e.}, without hardware noise) simulator. The subfigures (in order from a to e) are for uniform, binomial, triangle, bimodal, and arbitrary distributions. All QRNG subcircuits used five qubits, and the number of shots used for each simulation is given in Table \ref{table:qrng_shot_selection}. The blue histograms illustrate the desired PMF input to the QRNG subcircuit generation process, while the orange histograms illustrate the distribution measured when simulating the QRNG subcircuits.}
  \Description[QRNG Subcircuit Simulation Results]{Histograms illustrating the desired PMFs and the ideal (\textit{i.e.}, without hardware noise) simulated PMFs from QRNG subcircuits.}
\end{figure}

Using the computed shot counts, we then simulated QRNG subcircuits for a uniform, binomial, triangle, bimodal, and arbitrary distribution, as shown in subfigures a, b, c, d, and e of Figure \ref{fig:qrng_noiseless_simulation}.
Even a qualitative assessment of these results illustrates that using the shot counts computed with aid of the G-statistic seems to have effectively reduced sampling error: the orange and blue distributions are well-aligned, often to the point of being visually indistinguishable.
This demonstrates that \textit{MustangQ}'s QRNG subcircuits accurately represent the distributions they were designed for, thus providing an efficient way to prepare what are often non-trivial subcircuits.
For example, consider the binomial distribution, which is a parametric distribution with important consequences for post-quantum cryptography \cite{vadim12,ducas13}.
The binomial coefficients are relatively straightforward to compute: they are the scalar coefficients of the expanded polynomial, $(x+y)^N$, where $N$ is the number of qubits \cite{sinha23}.
However, even for a distribution with so much structure, the resulting circuit is non-trivial, requiring sixty-one gates, thirty-one of which are parameterized, meaning the unique parameterization values must be computed.
For even more complicated distributions, such as the arbitrary distribution, the subcircuit design is even more involved, and manual design is thus best avoided.

\begin{figure}[ht]
  \centering
  \includegraphics[width=0.9\linewidth]{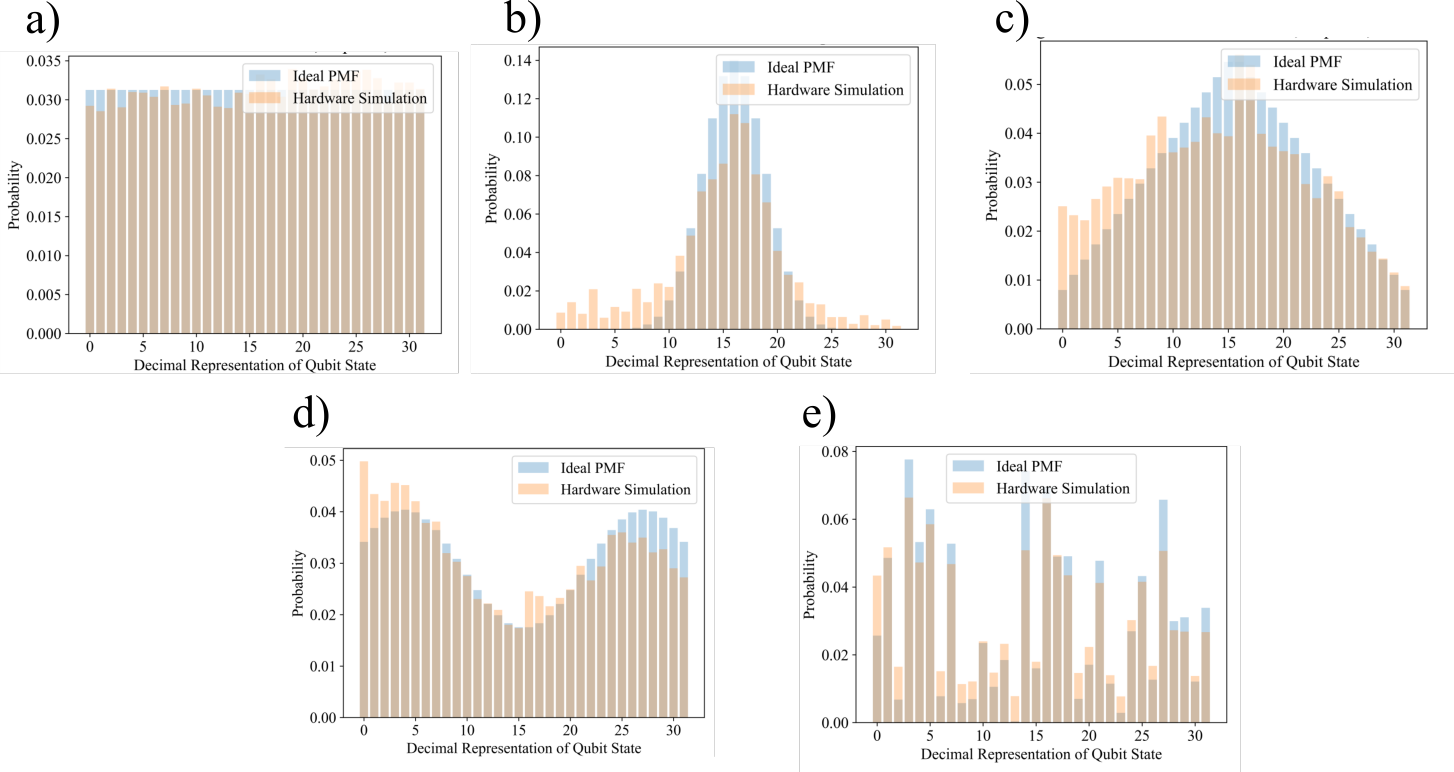}
  \caption{\label{fig:qrng_noisy_simulation}Results of simulating five QRNG subcircuits using the hardware simulator for IBM's Washington machine. The subfigures (in order from a to e) are for uniform, binomial, triangle, bimodal, and arbitrary distributions. All QRNG subcircuits used five qubits. The blue histograms illustrate the desired PMF input to the QRNG subcircuit generation process, while the orange histograms illustrate the distribution measured when simulating the QRNG subcircuits.} 
  \Description[QRNG Subcircuit Simulation Results]{Histograms illustrating the desired PMFs and the hardware-noise simulated PMFs from QRNG subcircuits.}
\end{figure}

Having considered how well the distributions resulting from simulated QRNG subcircuits align with the desired distribution in the presence of sampling noise, we performed a second set of experiments looking at the effects of hardware noise.
We used an IBM-provided simulator that mimics the noise signature of the Washington device, a 127-qubit machine that was---until recently---amongst its most sophisticated.\footnote{The Washington machine has been replaced by comparably-sized machines, including the Brisbane, Osaka, and Kyoto machines \cite{ibmq23b,ibmq23c}.}
Using the five qubits as selected by Qiskit's IBM transpiler, which uses qubit connectivity and performance information to make qubit mapping decisions, we obtained the results in Figure \ref{fig:qrng_noisy_simulation}.
Whereas the qualitative assessment of Figure \ref{fig:qrng_noiseless_simulation} suggested the QRNG subcircuit distributions accurately represented those they were designed to, Figure \ref{fig:qrng_noisy_simulation} suggests that hardware noise substantially reduces the ability of QRNG subcircuits to faithfully represent the desired PMFs.
Quantitative metrics are again in Appendix \ref{app:qrng_data}, and here we discuss three trends.

First, distributions without small values are more accurate than those with small values, even when the latter distributions are simpler.
For example, notice that the binomial and triangle QRNG subcircuit results have larger discrepancies from their ideal distributions than do the bimodal or arbitrary distributions.
This could be a consequence of the fact that smaller rotations required to prepare the smaller PMF distribution values lose precision, as discussed in Ref. \cite{sinha22} and in Sections \ref{sub:normalization} and \ref{sub:angle_encoding} of this work.
While the methods of Sections \ref{sub:normalization} and \ref{sub:angle_encoding} are not immediately applicable in the context of probability amplitude inputs, these results suggest that further research along similar lines might be important for circuit performance on near-term devices.

Second, circuits with lower circuit depth naturally perform better, emphasizing the need for robust optimization, and thus automated optimization toolkits such as \textit{MustangQ}.
For example, consider the uniform distribution, which has by far the lowest depth, and also by far the lowest G-statistic (and correspondingly high similarity metric).

Third and finally, we note a dramatic rise in the value of the G-statistic (and corresponding drop in the similarity metric) for the binomial distribution, which is interesting, as it is not amongst the most complex distributions considered, and the error does not appear more dramatic than in---for example---the triangle or bimodal distributions.
This substantial increase (or decrease, depending upon the quantity considered) is caused by the computational effect of events that occur in an experimental distribution but not in the theoretical distribution against which one is comparing.
Specifically, note that hardware noise caused the QRNG subcircuit to produce several events on the tails of the distribution that do not exist in the theoretical binomial distribution; this causes a dramatic effect in G-statistic value.
Alternative statistics---such as those discussed in Section \ref{sub:verification_evaluation}---that do not weight this discrepancy in the same way may provide a more intuitive assessment of the performance for a QRNG subcircuit that generates a binomial distribution \cite{sinha23}.

\begin{figure}[ht]
  \centering
  \includegraphics[width=0.9\linewidth]{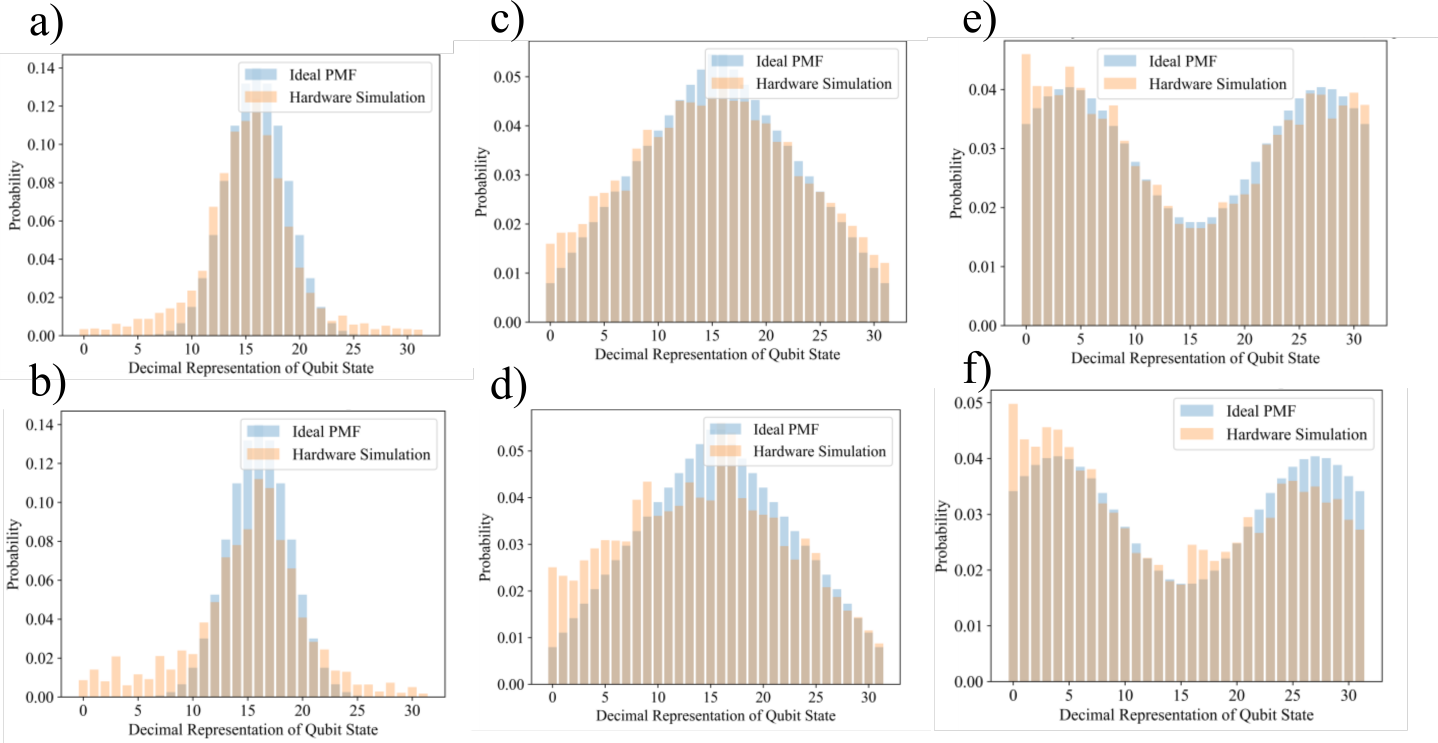}
  \caption{\label{fig:qrng_optimized_simulation}Results of simulating five QRNG subcircuits optimized as described in Section \ref{sub:symmetric_opt} using the hardware simulator for IBM's Washington machine. As in previous experiments, all QRNG subcircuits used five qubits, and the blue histograms illustrate the desired PMF input to the QRNG subcircuit generation process, while the orange histograms illustrate the distribution measured when simulating the QRNG subcircuits.} 
  \Description[Optimized QRNG Subcircuit Simulation Results]{Histograms illustrating the desired PMFs and the hardware-noise simulated PMFs from QRNG subcircuits that were optimized using an optimization that reduces gate count as described in Section \ref{sub:symmetric_opt}.}
\end{figure}

After assessing the performance of baseline QRNG subcircuits on simulated noisy hardware, we used the optimization of Section \ref{sub:symmetric_opt} to significantly reduce the gate count and depth of the QRNG subcircuits preparing binomial, triangle, and bimodal distributions.
These distributions allow for such optimization, because they are symmetric, and as described in Section \ref{sub:symmetric_opt}, the gate count and quantum depth can be reduced by approximately half for every line of symmetry in a PMF.\footnote{While the uniform distribution is also symmetric, its quantum circuit representation is so simple that it does not require further optimization.}
Figure \ref{fig:qrng_optimized_simulation} illustrates the results, and the qualitative improvement across all three distributions is clear.
(The top row shows the optimized results, while the bottom shows the unoptimized ones from Figure \ref{fig:qrng_noisy_simulation}.)
While the quantitative improvement is less clear for the binomial distribution (given the issue described above), the visual overlap is substantially better, particularly on the left and right tails.

\subsection{Quantum Oracles}\label{sub:MustangQ_oracles}
Oracles that preserve domain values cannot overwrite the values on input qubits with output values, and consequently, the ESOP synthesis approach is applicable, because it generates circuits that have one qubit for each input or output.
Specifically, for a classical function specified with domain bitstrings of length $n$ and range bitstrings of length $m$, the ESOP synthesis approach generates a circuit with $n+m$ qubits, in which qubits are initialized to the $n$ input values plus $m$ ancillary values of $\ket{0}$ and have outputs of the $n$ input values plus the now-evolved $m$ output values associated with the given inputs.
Therefore, \textit{MustangQ} synthesizes oracle subcircuits that preserve domain values with just two broad steps.
First, after receiving a valid \texttt{.pla} file as input, \textit{MustangQ} ensures that the function is represented as an ESOP by utilizing EXORCISM-4, a code written in the \texttt{C} programming language that produces not only an ESOP, but a minimized ESOP \cite{mishchenko01}.
Second, \textit{MustangQ} applies the ESOP synthesis method to the now-ESOP-formatted classical function to generate a circuit, as described in Section \ref{sub:esop_synthesis}.

For oracle subcircuits that utilize the minimal number of qubits, \textit{MustangQ} offers two approaches, both of which use TBS.
The most straightforward form of TBS---as described in Section \ref{sub:TBS}---requires the preprocessing steps of Sections \ref{sub:expand_assign_one_to_one} and \ref{sub:onto} be applied prior to the circuit synthesis itself (described in Section \ref{sub:TBS}).
As the remainder of this section will illustrate, the preprocessing required for TBS requires significant computation, which motivated us to explore an alternative TBS-based method that generates oracles with the minimal number of qubits but at lower computational cost.

Regardless of synthesis method, all oracle subcircuits can be optimized using the functionality described in Section \ref{sub:reduction_and_decomp}, and can be visualized or ported to circuits executable on the IBM suite of devices as described in \ref{sub:connecting_with_other_tools}.

To illustrate all of this functionality, we present two sets of results.
First, we compare oracle subcircuits synthesized from benchmark functions when using both the ESOP and TBS methods, including the variant we developed to reduce the computational cost of TBS.
Then---as in Section \ref{sub:MustangQ_QRNGs}---we illustrate \textit{MustangQ}'s ability to synthesize subcircuits that are both executable using quantum simulators and appropriate for inclusion in larger quantum circuits representing algorithms, such as Grover's search.

We begin with the benchmark functions: for twelve functions represented as \texttt{.pla} files, we synthesized oracle subcircuits using ESOP synthesis and TBS.
For each oracle subcircuit, Appendix \ref{app:oracle_data} reports quantitative data, while the remainder of this section discusses the two most significant trends.

\begin{figure}[ht]
  \centering
  \includegraphics[width=\linewidth]{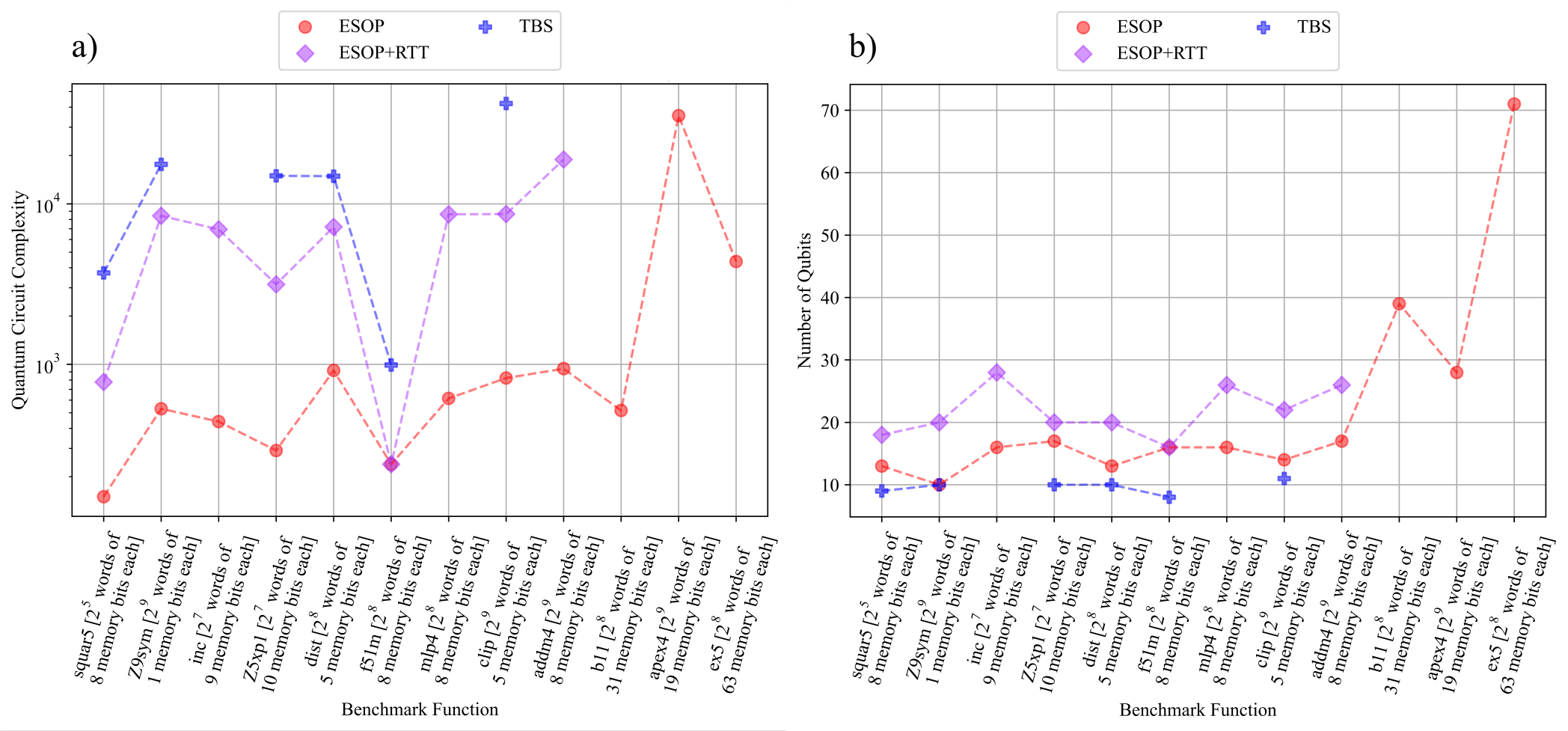}
  \caption{\label{fig:oracle_complexity_qubits}The circuit complexity (subfigure a) and qubit count (subfigure b) of twelve \texttt{.pla} functions when synthesized using ESOP synthesis (red circles), ESOP synthesis with non-required processing (purple diamonds), and TBS (blue crosses).  Some functions were too large to be expanded during RTT or to be addressed with TBS, which has an exponential algorithmic complexity; for these functions, \textit{MustangQ} timed out before completion, so no data point is shown.}
  \Description[Oracle Subcircuit Circuit Complexity and Qubit Count]{Two scatter plots illustrating the circuit complexity and qubit count of twelve oracle subcircuits generated with three synthesis methods each.}
\end{figure}

First, the ESOP synthesis and TBS methods provide different advantages: while ESOP nearly always produces circuits with lower complexity, TBS generally produces circuits with fewer qubits than ESOP synthesis.
(See Figure \ref{fig:oracle_complexity_qubits}.)
Both of these trends make sense in light of the synthesis methods' structures.
Specifically, TBS produces circuits with the minimal number of qubits given its `repurposing' of the qubits used to send in initial values.
For this set of benchmark functions, this qubit benefit manifests itself as an average reduction of over $56\%$ when comparing between TBS and ESOP synthesis.
However, the qubit benefit of TBS synthesis comes at the cost of being unable to leverage incompletely-specified functions.
While TBS as described in Section \ref{sub:TBS} requires preprocessing that eliminates the dashes present in \texttt{.pla} files, ESOP synthesis can leverage these unspecified values to reduce the number of controls on its Toffoli gates.
Specifically, variables specified as either a $0$ or a $1$ result in a cost of either $3$ (for two Pauli-$\mathbf{X}$ gates and a control) or $1$ (for a control), respectively, but variables that are specified as dashes correspond to \textit{no} controls---and thus no cost.
To further support our hypothesis that some of the difference between the average circuit complexities of ESOP and TBS synthesis is a consequence of leveraging incompletely-specified functions, we use \textit{MustangQ} to produce oracles with a third, non-practical method: ESOP synthesis after having fully preprocessed the classical function specifications, as if we would run TBS.
As illustrated in subfigure a of Figure \ref{fig:oracle_complexity_qubits}, the fact that ESOP synthesis with preprocessing has a higher average complexity than ESOP synthesis without ($6.981.6$ versus $3,769.5$) further supports the conclusion that ESOP synthesis is able to substantially reduce cost by utilizing incompletely-specified functions.

\begin{figure}[ht]
  \centering
  \includegraphics[width=0.5\linewidth]{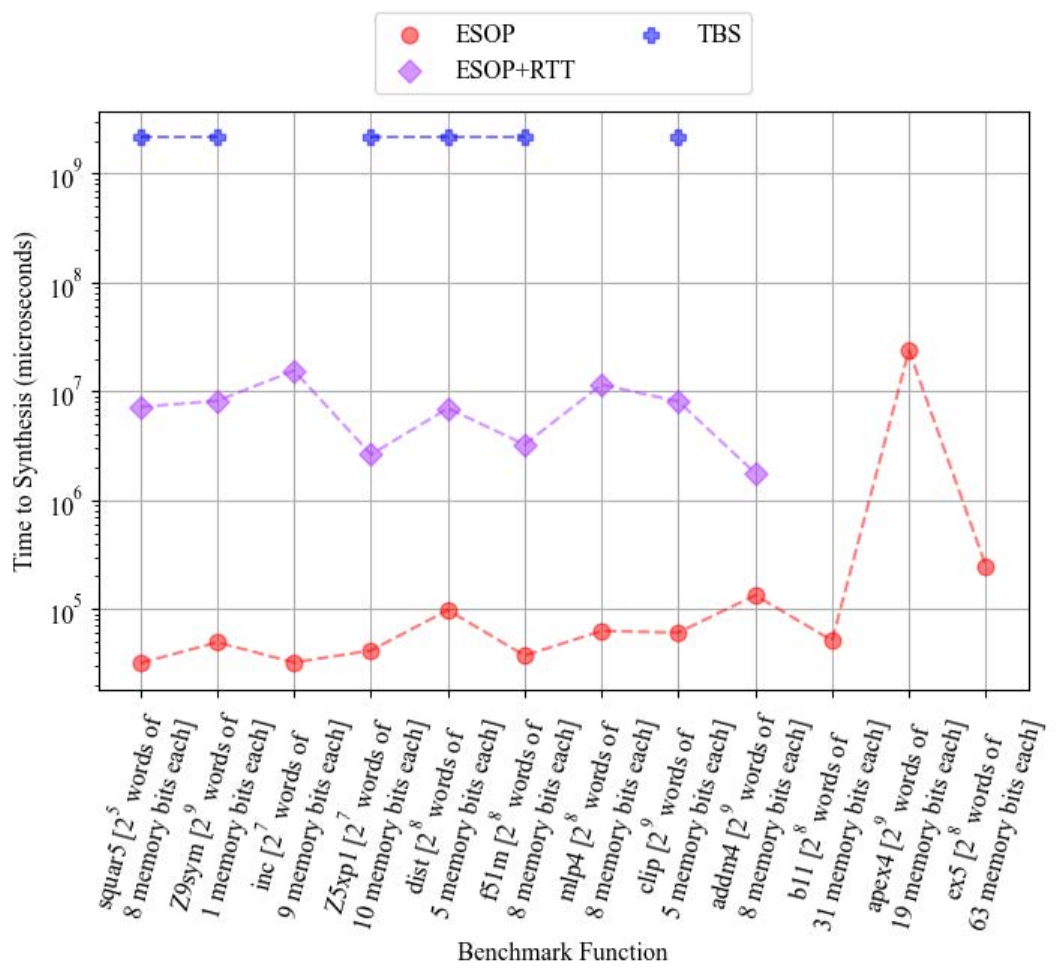}
  \caption{\label{fig:oracle_time}The time-to-synthesis (in microseconds) of twelve \texttt{.pla} functions when synthesized using ESOP synthesis (red circles), ESOP synthesis with non-required processing (purple diamonds), and TBS (blue crosses).  Some functions were too large to be expanded during RTT or to be addressed with TBS, which has an exponential complexity; for these functions, \textit{MustangQ} timed out before completion, so no data point is shown.}
  \Description[Oracle Subcircuit Time-to-Synthesis]{A scatter plot illustrating the time-to-synthesis of twelve oracle subcircuits generated with three methods each.}
\end{figure}

The second noteworthy trend is that---while ESOP synthesis and TBS offer differing strengths when it comes to the circuits produced---obtaining an oracle with the minimal number of qubits comes at a significant temporal cost.
As illustrated in Figure \ref{fig:oracle_time}, ESOP synthesis was not only faster than either ESOP synthesis with processing and TBS, but it was able to run to completion for all functions considered, while neither ESOP synthesis with preprocessing or TBS could do the same.
This again makes sense given the unavoidable complexity of the preprocessing required by TBS (and utilized to make a point with ESOP plus preprocessing): the cost of making the \texttt{.pla} functions one-to-one and onto is $O(2^{(n+m)})$, where $n$ and $m$ are the numbers of bits in the input and output bitstrings, respectively.
Consequently, TBS is not scalable to the degree that we want \textit{MustangQ}'s synthesis methods to be, even as we do not discount its benefit of minimal qubit usage.

TBS' costs illustrated by \textit{MustangQ} and its important minimal qubit usage led us to explore TBS and to develop several variants, one of which is described in Section \ref{sub:TBS_RM}.
Because the TBS-RM method considers input/output pairs that have already been synthesized, we hypothesized that it would be able to reduce the circuit complexity of the oracle subcircuits while still using the minimal number of qubits.
We found that this was indeed the case, although the cost benefits were limited: the circuit complexity improved by an average of $980$, which is a fraction of the complexities of these circuits.
Furthermore, the time complexity issues were further exacerbated; one of the oracle subcircuits that we could obtain with the basic TBS method was unobtainable in this form (\texttt{clip}).
Consequently, our investigation into more efficient synthesis---in terms of circuit complexity and time-to-synthesis---is ongoing, and we are working to improve TBS' preprocessing stages to reduce these metrics.
Specifically, we are investigating alternative assignments of the output `don't cares,' such that they provide functions that are better suited to the identity-transformation procedure that is the core of TBS.

To close this section, we consider synthesis of smaller oracle subcircuits that could be run on the IBM suite of quantum devices.\footnote{These examples require only seven qubits, so could be run on several of IBM's existing machines. However---as mentioned in Section \ref{sec:introduction}---these experiments were run using IBM's quantum simulator given limited hardware access.}
This achieves our dual goal of illustrating \textit{MustangQ}'s scalability using large benchmark circuits, while also---as in Section \ref{sub:MustangQ_QRNGs}---illustrating its applicability to near-term hardware with smaller, tailored examples.
We consider synthesizing oracle subcircuits for Grover's search algorithm, in which the goal is to search a `deck of cards.'
We can represent a fifty-two card deck with six bits: two for the suit, and four for the card value.
For example, suppose the diamonds suit is represented as $10$; then the ten of diamonds would be represented as $101010$.
Similarly, suppose the clubs suit is represented as $00$; then the two of clubs would be represented as $000010$.
This system results in fifty-two bitstrings that represent playing cards and twelve `unused' bitstrings, a consequence of the fact that fifty-two is not a power of two.

\begin{figure}[ht]
  \centering
  \includegraphics[width=\linewidth]{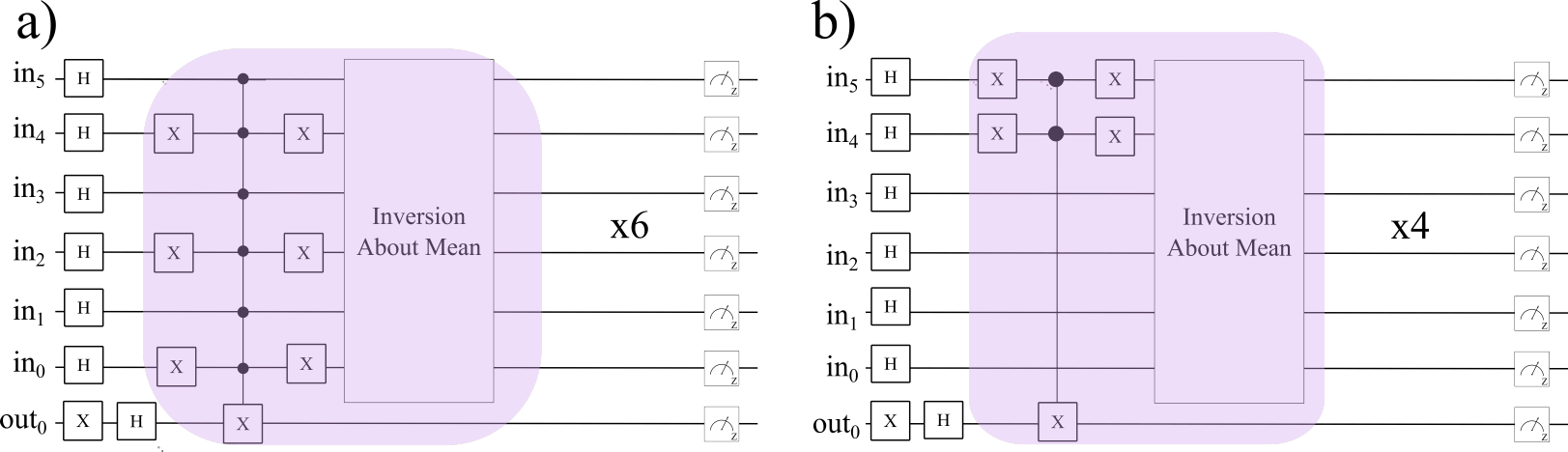}
  \caption{\label{fig:grover_card_example_circuits}Two circuits for implementing Grover's search algorithm; subfigure a searches for ten of diamonds, encoded as 101010, and subfigure b searches for any card with a clubs suit, meaning first two bits 00. The oracle subcircuits are the first of the two components highlighted in purple (\textit{i.e.}, everything except the ``Inversion About Mean.''}
  \Description[Grover's search for a Deck of Cards]{Two Grover's search circuits containing the oracle subcircuits generated by \textit{MustangQ}; these circuits contain the initialization portion of Grover's Algorithm, followed by the oracles, followed by the standardized inversion about the mean unitary, which depends solely on the size of the database.}  
\end{figure}

Grover's search algorithm has three portions, the first and third of which are standardized given the size of the database to be searched.
\textit{MustangQ}'s role is to synthesize the non-standard oracle subcircuits, and this section presents two examples: searching for a single card (the ten of diamonds, represented as $101010$) and for all cards with a suit of clubs (represented with the first two qubits $00$).
Figure \ref{fig:grover_card_example_circuits} illustrates the oracle subcircuits (the gates highlighted in purple) in the context of Grover's algorithm.
In both cases, the oracles are very simple, given the small database size and conceptual simplicity of these examples.
And yet, the fact that they differ highlights the utility of a toolkit like \textit{MustangQ}: developing circuits for far larger differing functions quickly makes repeated oracle subcircuit synthesis prohibitively difficult.

\begin{figure}[ht]
  \centering
  \includegraphics[width=\linewidth]{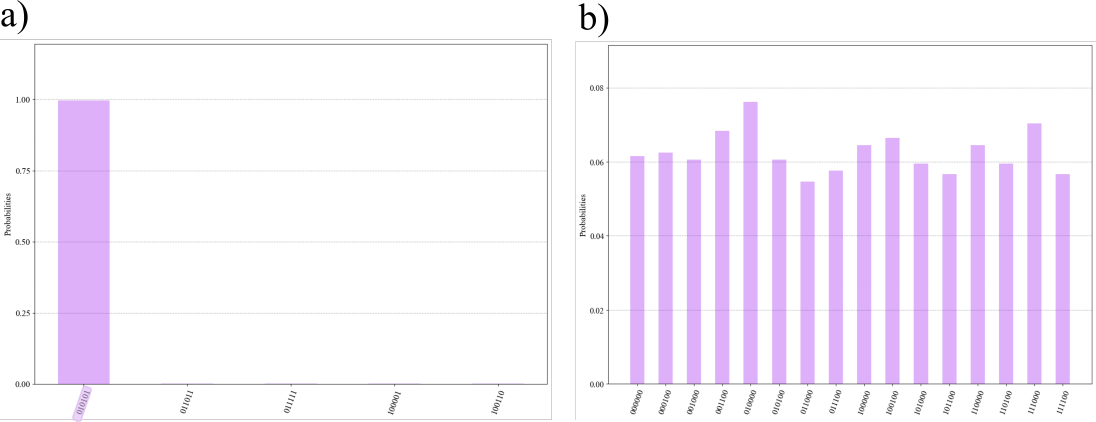}
  \caption{\label{fig:grover_card_example_results}Two sets of Qiskit simulation results for the Grover's search circuits of Figure \ref{fig:grover_card_example_circuits}, run using 1024 shots. For subfigure a (searching for the 10 of diamonds), the correct result was found in 99\% of the 1024 shots; note that Qiskit lists qubits in an order \textit{opposite} that expected, meaning that 101010 appears as 010101 on the $x$-axis. For subfigure b (searching for all cards of the club suit), we measure sixteen different bitstring encodings, each of which is either the Ace through King of clubs or one of the three `unused' numerical values (0000, 1110, 1111) for the clubs suit (meaning with $00$ as the first two bitstring values). (Again, because of Qiskit's qubit ordering, the values on the $x$-axis have $00$ as their \textit{last} two qubit values.}
  \Description[Simulation Results for Two Circuits Implementing Grover's Search Algorithm]{Two sets of results for Qiskit simulations running the Grover's search cirucits of Figure \ref{fig:grover_card_example_circuits}; the first has a single visible bar, indicating that the vast majority (nearly 100\% of bitstrings) were the ten of diamonds bitstring.  The second has sixteen visible bars, each of which represents approximately 6\% of cases, and each of which has a suit bitstring of $00$, for clubs.}
\end{figure}

As indicated in Figure \ref{fig:grover_card_example_circuits}, the highlighted portions of the circuits are repeated several times, increasing the likelihood that---upon measurement---the  circuit's qubits will collapse to a desired state.
It can be shown that the number of repetitions required is $O(\sqrt{N})$, where $N$ is the number of elements in the search space, and this provides the speedup over classical methods, which have a complexity $O(N)$ \cite{nielsen11}.
Figure \ref{fig:grover_card_example_results} illustrates the results after using Qiskit's ideal simulator to simulate running these circuits on quantum hardware; subfigure a shows correctly obtaining the ten of diamonds state (101010) in 99\% of 1024 individual circuit executions,\footnote{IBM---and several other commercial quantum hardware providers---term individual circuit executions ``shots.''} while subfigure b shows correctly obtaining \textit{only} states that have the first two bitstring values $00$.

It is worth noting one more aspect of these results; although the number of oracle/iteration about the mean repetitions recommended depends upon the size of the database, we repeated these portions of the circuit differently for two search queries that have the same database size.
We determined these iteration counts experimentally, because we found that---particularly for small databases---one cannot blindly apply the `recommendation' of $O(\sqrt{\frac{N}{M}})$ iterations, where $N$ is the number of items in the database, and $M$ is the number of database items that are solutions to the search query.
This is a feature of Grover's search algorithm that does not seem to be oft-recognized in other resources.
While it is is widely recognized that Grover's algorithm is an iterative one---meaning a single application of an oracle and inversion about the mean is likely insufficient for obtaining accurate results---the fact that applying too many iterations can generate results that are equally poor in quality to those derived from a circuit with too few iterations seems to be often unstated.
Many sources present the mathematics from which this consequence can be realized, but no sources that we found carried out that explicit series of steps.
The only source that we found to qualitatively state this feature at all is a Medium article \cite{farina2021}.
Thus, in Appendix \ref{app:grover_exploration}, we briefly discuss the oscillatory nature of Grover's search algorithm by considering both its geometric presentation (as described in Ref. \cite{nielsen11}) and experimental results from our deck of cards examples.
While our primary goal of that discussion is to describe an interesting---yet seemingly little-discussed---feature of a seminal quantum algorithm, a secondary goal is to illustrate, in an admittedly small way, the sort of research that is possibly of more interest than construction of circuits, and that we believe toolkits like \textit{MustangQ} make more accessible.

\section{Conclusion and Future Work}\label{sec:conclusion}
In this work, we describe \textit{MustangQ}, a toolkit for automating synthesis, optimization, and production of quantum circuits from classically-specified functions.
In addition to describing its core functionality, we illustrate its application to three types of quantum subcircuit.
Quantum subcircuits make an ideal case for automation of circuit synthesis and optimization, because they must be synthesized and optimized for each possible situation, so it is not feasible to manually produce optimized subcircuits at scale.
This paper shows how \textit{MustangQ} provides a consistent framework with a variety of tools to address this scalability issue, automating everything from classical function preprocessing to eventual mapping onto commercially-existing hardware via connection with other available quantum software tools.
Furthermore, this work illustrates how the toolkit enables research discoveries and deeper understanding of even well-studied quantum algorithms. 

While \textit{MustangQ}'s use thus far in promoting research findings is an exciting and promising sign of its utility, development is far from over; we have several plans to make its current abilities more robust and to expand those abilities via future research.
On the development side, we plan to add further alternative input and preprocessing methods (requiring new parsers and potential custom data structures), quantum synthesis methods (including methods based on function slicing and re-alignment), quantum optimizations (including technology-independent rotation-based optimization), and enhanced verification (including more robust interfacing with the QMDD structure).

For future research, we have several goals well-suited to \textit{MustangQ}'s flexible yet consistent framework, including exploring the gate count, quantum cost, and quantum depth ramifications of different classical preprocessing methods; deriving additional synthesis methods that do not require expensive function preprocessing; investigating how the benefits of different data encoding methods might translate to benefits in quantum algorithms that currently use less-than-ideal encoding methods; and further exploration of circuit performance---and ways to improve it---for available quantum hardware.
Hence, there is no shortage of good work to be done, and we look forward to extending both the capabilities of the \textit{MustangQ} and what can be learned using it.

While many lessons to be learned from \textit{MustangQ} are most relevant for circuits that remain too large and complex for contemporary machines, that is not likely to be true for long.
Rapid improvements in quantum hardware motivate development of quantum synthesis software that will make the future of quantum computing as rich as its classical counterpart is today.

\section{Acknowledgements}
We gratefully acknowledge Dr. D. Michael Miller for helpful discussions and provided software.

\bibliographystyle{ACM-Reference-Format}
\bibliography{references}

\clearpage

\appendix

\section{Appendix: Preprocessing Classical Functions}\label{app:preprocessing_details}
\subsubsection{Expanding, Assigning, and Making One-to-One Classical Specifications}\label{sub:expand_assign_one_to_one}
As described in Section \ref{sub:pla_format}, the \texttt{.pla} format reduces the size of classical function specifications, which is generally a desirable feature.
However, some quantum circuit synthesis approaches require that a function be fully-specified, meaning that functions represented by \texttt{.pla} files must be expanded to remove the multi-valued dashes.
Additionally, some quantum synthesis methods require that classical function specifications be one-to-one functions prior to synthesis.
Consequently, \textit{MustangQ} can embed classical specifications into functions that preserve the domain to range mappings while meeting these criteria.

First, the tool automates function expansion, which is the process of removing dashes on the length-$n$ bitstring variable valuations: each variable valuation that contains a dash is replaced with $2$ bitstrings, one in which the dash is evaluated as a $0$, and another in which the dash is evaluated as a $1$.
For variable valuations containing multiple dashes, this procedure occurs for each, meaning a bitstring with $q$ dashes is converted into $2^q$ separate bitstrings.
It is again worth emphasizing that this step is not uniformly required to synthesize quantum circuits using \textit{MustangQ}, but is rather a condition for synthesizing using some approaches, such as the most basic form of TBS.

Second, \textit{MustangQ} automates function assignment, in which dashes are removed from the length-$m$ bitstring function evaluations.
Here, the dashes denote ``don't cares,'' meaning that the bit replaced with a dash does not matter to the function and may have either a $0$ or $1$ value.
\textit{MustangQ} currently replaces these dashes with $0$ values during expansion, and we are exploring synthesis methods that allow for assigning output don't cares heuristically and according to synthesis-algorithm-specific constraints so as to reduce circuit complexity.

Third and finally is the process of making functions one-to-one; there are several methods for embedding a function of interest into a one-to-one representation \cite{miller09}.
Currently, \textit{MustangQ} implements a three-step process termed RTT \cite{gabrielson18}, which embeds functions of interest into one-to-one functions that will use the minimal number of qubits.
First, RTT determines the maximum number of times that any $m$-bit function valuation is duplicated.
The maximum number of duplicates is denoted $N_{dup}$, and if $N_{dup}=0$, then the function of interest is already one-to-one, and the RTT process is finished.
Otherwise, RTT moves to the second step of adding ancilla inputs and garbage outputs to the function domain/range pairs that have duplicated range values.
(It is worth emphasizing that the reversibility requirement of quantum circuits requires that we not only add garbage values to the outputs, but also that we add ancilla to the inputs associated with duplicated outputs.)
The number of garbage outputs to be added is given by $v = \log_2 (N_{dup})$, and the number of ancilla inputs to be added is given by $w = v + m - n$.
All new garbage and ancilla values are initialized to the value zero.
Third and finally, RTT sets the ancilla and garbage values, such that every input/output pair from the function of interest is now embedded in a one-to-one representation.
For each group of $m$-bit function-value duplicates, the first duplicate instance's garbage remains assigned to $0$, and each subsequent duplicate instance is assigned to the value of the previous instance incremented by $1$, and the 
same process is used to address the $n$-bit function values associated with the $m$-bit function valuations.

\subsubsection{Making Onto Classical Specifications}\label{sub:onto}
Some quantum circuit synthesis methods require classical function specifications that are not only one-to-one, but which are also onto.
(Again, other synthesis methods require neither of these steps; this section simply discusses the available processing steps in \textit{MustangQ}.)
As with approaches for making functions one-to-one, there are several methods for embedding functions of interest into onto functions, and \textit{MustangQ} implements two of the methods described in Ref. \cite{miller09}.

One approach is a straightforward one that randomly assigns unassigned domain values to as-yet-unused range values.
All one-to-one bitstring-based representations will definitionally have the same number of as-of-yet unassigned domain and range values, so we may simply pair the first domain value to not have a specified range value with the first range value to not have a specified domain value until no such values exist.

A second approach requires more computational time, but benefits at least one circuit synthesis method: TBS generally produces better-optimized circuits when as many domain values are as close to their corresponding range values as possible.
We thus implemented a second approach to automating onto-function creation by pairing as many unspecified domain values as possible to identical---or nearly identical---range values.
The process has two steps: first, as many unassigned domain values as possible are matched with identical unused range values.
Then, once all possible matching domain/range pairs have been set, each remaining domain value is assigned an as-of-yet unused range value that is the lowest possible Hamming distance away.
The second step is perhaps best illustrated with an example: consider a function with $n=m=3$ for which $001$ is an unassigned domain value that could be paired with either of two as-of-yet unused range values, $110$ or $011$.
Here, $001$ would be paired with $011$, because those values have a Hamming distance of $1$, while $001$ and $110$ have a Hamming distance of $3$.

\subsubsection{Normalizing Classical Specifications}\label{sub:normalization}
Several quantum synthesis algorithms require the classical input specifications to be normalized, such that all function valuations fall within an interval determined by the situation, or such that all the values in a specification sum to one.

First, we consider normalization that keeps all values within an interval.
Usually, that interval is $[0, 2\pi)$, because the synthesis methods that require normalization usually apply rotation gates, which act to rotate qubits modeled as points on the Bloch sphere by specified angles about the $x$-, $y$-, or $z$-axes.\footnote{Alternatively, one might choose a normalization range of $[-\pi,\pi)$ or $(-\pi,\pi]$ to accommodate signed values. For the remainder of this section, we consider a range of $[0, 2\pi)$ while acknowledging that other ranges would work equally well.}
There are several ways to normalize, and different approaches are preferable in different situations.
This section describes three normalization approaches, including a novel approach that provides improved value precision when representing values as bitstrings.

First is the most straightforward approach: normalize each function value to one within the range $[0, 2\pi)$ by determining the maximum value in the function specification, $v_{max}$, and multiplying every function value by a normalization factor, $f_{norm} = \frac{2\pi}{v_{max}}$.
For synthesis methods utilizing both the $\theta$ and $\phi$ angles of a parameterized qubit, separate $V_{max}$ and $f_{norm}$ values are required.

A second approach assumes that every classically-specified value is a positive, fixed-point value, in which the position of the fixed-point determines the interval into which the data falls.
Each classical bitstring value is interpreted as
\begin{math}
    0.b_{-1}b_{-2} \cdots b_{-m+3}b_{-m+2} \text{, where } b_j \in \mathbb{Z}_2
\end{math}, which allows for values within 
\begin{math} 
[0, 1) \in [0, 2\pi)
\end{math}.
This eliminates the need to compute $V_{max}$ and perform $N=2^n$ multiplication operations but at the cost of using only a small portion of the available $[0, 2\pi)$ interval.
Two variants of this approach thus better utilize that range.
A first naive modification is to multiply every interpreted value by $f_{norm} = 2\pi$, thus utilizing the full $[0, 2\pi)$ interval while avoiding computation of $V_{max}$ values, but at the cost of re-introducing $N=2^n$ operations.
Consequently, a more sophisticated variation is to adjust how the fixed-point value is interpreted.
If we assume that the fixed-point value has the form 
\begin{math}
    b_1b_0.b_{-1}b_{-2} \cdots b_{-m+3}b_{-m+2}
\end{math}, then all possible memory word values fall within the range $[0, 4)$, which still incurs loss of precision from the $[0, 2\pi)$ interval, but less than the reduction from $[0, 2\pi)$ to $[0, 1)$.
Furthermore, it allows for increased precision without either the $N=2^n$ multiplications or the $V_{max}$ computations.

A third normalization approach provides better precision for classical function specifications that have a large dynamic range, meaning the difference between the largest and the smallest function valuation is very large \cite{sinha22}.
Normalization that involves multiplication by $f_{norm}$ can result in fixed-point values with a large number of leading zeros, which cause the loss of significant bits due to the finite bitstring length used.\footnote{It is worth noting that, theoretically, quantum circuits can represent values with infinite precision when they are encoded using $(\theta, \phi) \in \mathbb{R}$. However, when the data to be represented comes from classical memory image files with discrete and restricted resolutions, the precision of the quantum circuit is inevitably limited.}
Consequently, we seek a normalization method that avoids such loss of precision for small-magnitude values, a problem that worsens as the dynamic range of a given classical function grows and as the number of bits used to represent the function values ($m$) decreases.

One partial solution would be to use floating-point arithmetic, instead of fixed-point arithmetic, during the normalization process; however, this introduces additional issues given conversion between fixed- and floating-point representations, as well as significant performance penalties arising from use of floating-point versus fixed-point arithmetic.
Therefore, a more robust approach (originally described in Ref. \cite{sinha22}) \textit{interprets} values in a way analogous to floating-point representations in conventional digital systems; each $j^\text{th}$ function range value is represented as a significand and an exponent, $S_j\times2^{E_j}$.
Such a representation can be obtained in four steps.

First, determine the number of leading zeros for every function range value, and denote this $z_j$ for the $j^\text{th}$ memory word.
Second, determine the largest $z_j$, $z_{max}=max\{z_j\}$, which will be an upper bound of possible exponent values, $E_j \in \{0, 1, \cdots, z_{max}\}=\mathbb{Z}_{z_{max}}^+$.
Third, interpret the significand of each function range value, $S_j$, as a fixed-point number that is left-shifted by $z_j+2$ bits and right-padded with $m-(z_j+2)$ zeros, where $m$ is the number of bits in each function range value.
Each of these $S_j$ values is then interpreted as having a fixed-point between the second and third bits from the leftmost side of the shifted bit string such that $S_j \in [0,4)$, since its left-shifted form is 
\begin{math}
    S_j=(b_1b_0.b_{-1}b_{-2} \cdots b_{m-z_j}00\cdots0)
\end{math},
where $b_1=1$ due to the shifting operation.
Fourth, we determine the exponent values, $E_j$, which also need to be normalized.
There are several approaches for such normalization, such as applying the normalization factor approach, but this would have all of the downsides discussed above.
An alternative is to leverage the fact that $E \in \mathbb{Z}_{z_{max}}^+$ is an integer: the real-valued interval $[0, 2\pi)$ could be partitioned into $z_{max}$ sub-intervals, and a given valuation of $E$ could be assigned to any value within the appropriate sub-interval.
While this option allows for optimization of the eventual quantum subcircuit, it can also complicate retrieving the original data from circuit measurement.
Consequently, another option implements a `round-to-nearest integer' rule that recreates the integer-valued exponent, $E$; this method offers the added advantage of `self-correcting' small inaccuracies introduced in the $E$ value given contemporary, non-ideal quantum gates.

Yet another variant for further increasing precision via value interpretation, and not computation, is to use a ``hidden bit'' \cite{ieee19}.
When using a hidden bit, the most significant bit of the significand is always set to either $0$ or $1$, such that tools interpreting the values can include an additional least significant bit in the significand.

A second type of normalization ensures that a set of variables sums to one, which is useful when such a set of values is to be considered a set of probabilities, as in Section \ref{sub:amplitude_encoding}.
This requires dividing each value in the set, $x_i$, by the square root of the sum of values in the set, $\sqrt{\sum_{j=1}^N x_j^2}$.

\section{Appendix: \textit{MustangQ}-Inspired Exploration with Grover's Algorithm}\label{app:grover_exploration}
The goal of this section is to explore a feature of Grover's algorithm that does not seem to be oft-discussed; as mentioned in Section \ref{sub:MustangQ_oracles}, we found only one source that qualitatively noted the cyclic accuracy/iteration relationship \cite{farina2021}.
Here, then, we describe the phenomenon in detail, including by showing the experiments with \textit{MustangQ}-generated oracles that first made us aware of it.
We utilize the notation of Ref. \cite{nielsen11}.
We begin with the geometric presentation of Grover's algorithm, which begins with two states, $\ket{\alpha}$ and $\ket{\beta}$:

\begin{equation}
\label{eq:alpha}
\ket{\alpha} = \frac{1}{\sqrt{N-M}}\sum_{x}^{''}\ket{x}
\end{equation}

\begin{equation}
\label{eq:beta}
\ket{\beta} = \frac{1}{\sqrt{M}}\sum_{x}^{'}\ket{x}\end{equation}

Here, $\sum_{x}^{'}\ket{x}$ denotes the summation of states that are solutions to the search problem, and $\sum_{x}^{''}\ket{x}$ denotes the summation of states that are not.
The initial state of a circuit prepared for Grover's search can then be represented as a linear combination of $\ket{\alpha}$ and $\ket{\beta}$ with equal probabilities for all results, given the Hadamard gates that put all qubits into a uniform superposition.
Specifically, the initial state $\ket{\psi}$ is given by,

\begin{equation}
\label{eq:initial_grover_state}
\ket{\psi} = \frac{1}{\sqrt{N}}\sum_{x=0}^{N-1}\ket{x} 
= \sqrt{\frac{N-M}{N}}\sqrt{\frac{1}{N-M}}\sum_{x}^{''}\ket{x} + \sqrt{\frac{M}{N}}\sqrt{\frac{1}{M}}\sum_{x}^{'}\ket{x}
= \sqrt{\frac{N-M}{N}}\ket{\alpha} + \sqrt{\frac{M}{N}}\ket{\beta}
\end{equation}

Geometrically, the goal of Grover's algorithm is to adjust $\ket{\psi}$ such that it aligns as closely as possible with $\ket{\beta}$.
Such alignment is achieved by iteratively applying the oracle and inversion about the mean subcircuits, each of which applies a reflection to the $\ket{\psi}$ state \cite{nielsen11}.
In particular, the oracle operation applies a reflection about $\ket{\alpha}$, and the inversion about the mean applies a reflection about $\psi$ \cite{nielsen11}.
As the product of two reflections is a rotation, each oracle/inversion about the mean operation is a rotation of the circuit state $\ket{\psi}$ \cite{nielsen11}.

\begin{figure}[ht]
  \centering
  \includegraphics[width=0.5\linewidth]{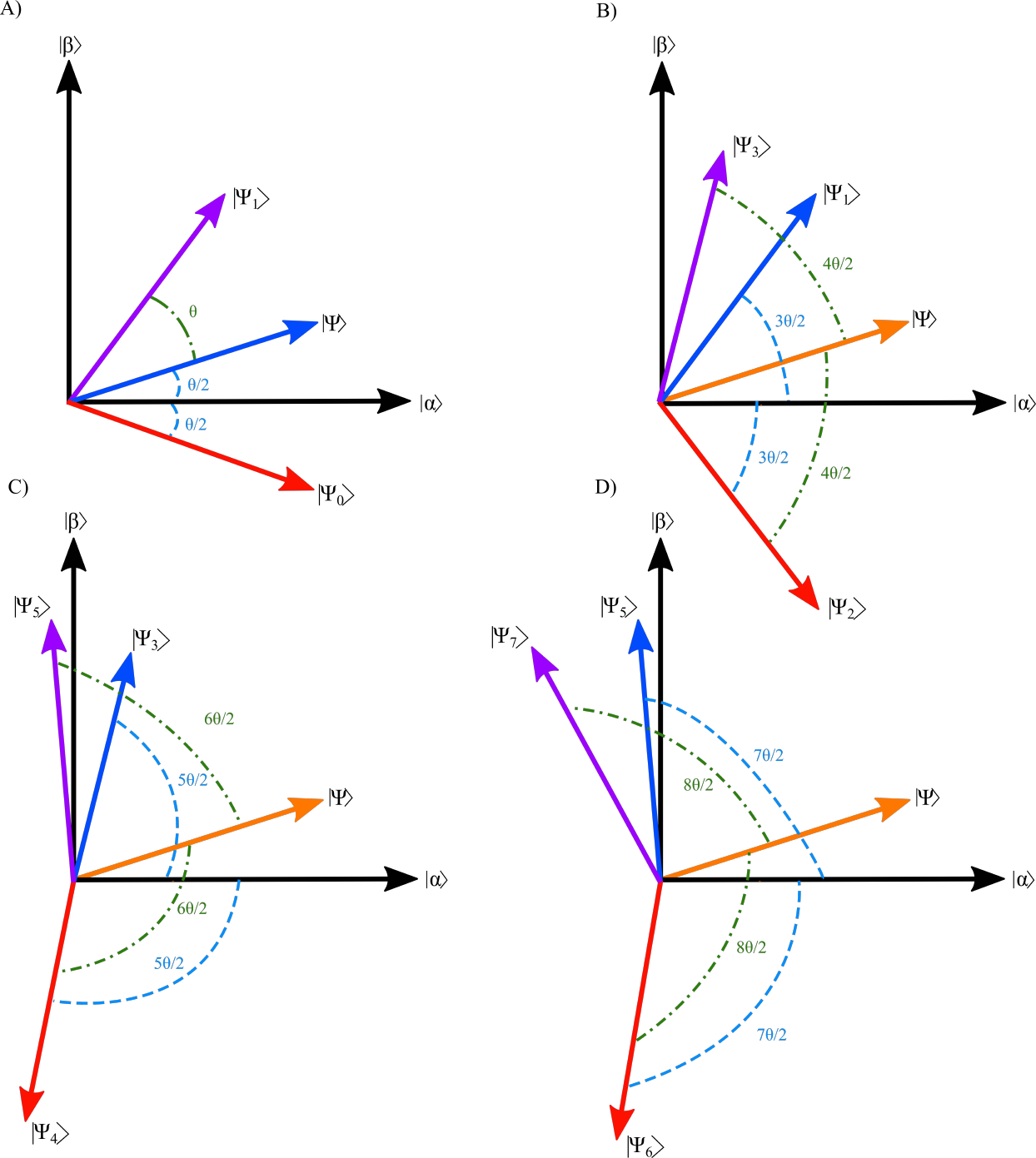}
  \caption{\label{fig:grover_geometry_illustration}Four applications of a Grover iteration (\textit{i.e.}, the oracle and inversion about the mean steps) generate the vectors marked in purple at each step.  Thus, the best time to measure illustrated here would be after the third iteration, when $\ket{\psi_5}$ is closer to $\ket{\beta}$ than $\ket{\psi_1}$,$\ket{\psi_3}$, or $\ket{\psi_7}$.  The blue vector in each cartoon is the initial state for the current iteration, and in all iterations except for the first, the initial state, $\ket{\psi}$, is denoted with an orange vector, such that it is straightforward to perform the inversion about the mean reflection over $\ket{\psi}$.}
  \Description[Geometrically illustrating four iterations of Grover's search algorithm.]{A four-part illustration of four iterations of Grover's search algorithm, which shows how too many iterations can move the solution state vector past the desired stopping point, just as too few can prevent it from reaching that stopping point. The figures show how the vectors move counterclockwise during a series of rotations and reflections.}
\end{figure}

This becomes more clear when re-writing $\ket{\psi}$ in terms of cosine and sine of an angle, $\theta$, such that $\cos({\theta/2})$ is defined to be equal to the coefficient on $\ket{\alpha}$, meaning $\cos{\frac{\theta}{2}} = \sqrt{\frac{N-M}{N}}$ \cite{nielsen11}.
We thus define $\theta$ in terms of the number of items in a database and the number of items that are solutions to the search query.
This definition makes $sin{\frac{\theta}{2}}=\sqrt{\frac{M}{N}}$, meaning we may write the initial state from Eq. \eqref{eq:initial_grover_state} as,

\begin{equation}
\label{eq:initial_grover_state_theta}
\ket{\psi} = \cos{\frac{\theta}{2}}\ket{\alpha} + \sin{\frac{\theta}2}\ket{\beta}.
\end{equation}
The two reflections are illustrated in subfigure a in Figure \ref{fig:grover_geometry_illustration}, which is based on Figure 6.3 from Ref. \cite{nielsen11}.
As described there, the initial vector, $\ket{\psi}$ in blue, is first reflected about $\ket{\alpha}$, giving the vector $\ket{\psi_0}$ in red.
Then, that vector is reflected over $\ket{\psi}$, giving the vector $\ket{\psi_1}$ in purple.
Consequently, a single application of the Grover oracle and inversion about the mean changes the state of the circuit from that in Eq. \eqref{eq:initial_grover_state_theta} to,

\begin{equation}
\label{eq:modified_grover_state_theta}
G\ket{\psi} = \cos{\frac{3\theta}{2}}\ket{\alpha} + \sin{\frac{3\theta}2}\ket{\beta}.
\end{equation}
Thus, one Grover iteration is a net rotation of the starting state by $\theta$ radians \cite{nielsen11}.

This illustrates why it is possible both to have too few and too many iterations of the Grover oracle and inversion about the mean.
Consider subfigure b, which illustrates what has happened in our example after a second iteration.
If we chose to stop and measure at that point, we would be more likely to measure at least one of the states in $\ket{\beta}$, given state $\ket{\psi_3}$'s similarity to that linear combination of desired states.
However, without measuring our state, we cannot know---absent working out all of the mathematics, which is precisely what we want the quantum computation to handle for us---how close $\ket{\psi_3}$ is to the desired state.
So, if we instead performed another iteration---as illustrated in subfigure c---the quantum state $\ket{\psi_5}$ would be even closer to $\ket{\beta}$.
However, we note that it has also moved \textit{beyond} $\ket{\beta}$, such that a fourth iteration---illustrated in subfigure d---would generate a state that is nearly as far from the desired state as that from the first iteration.

\begin{figure}[ht]
  \centering
  \includegraphics[width=0.8\linewidth]{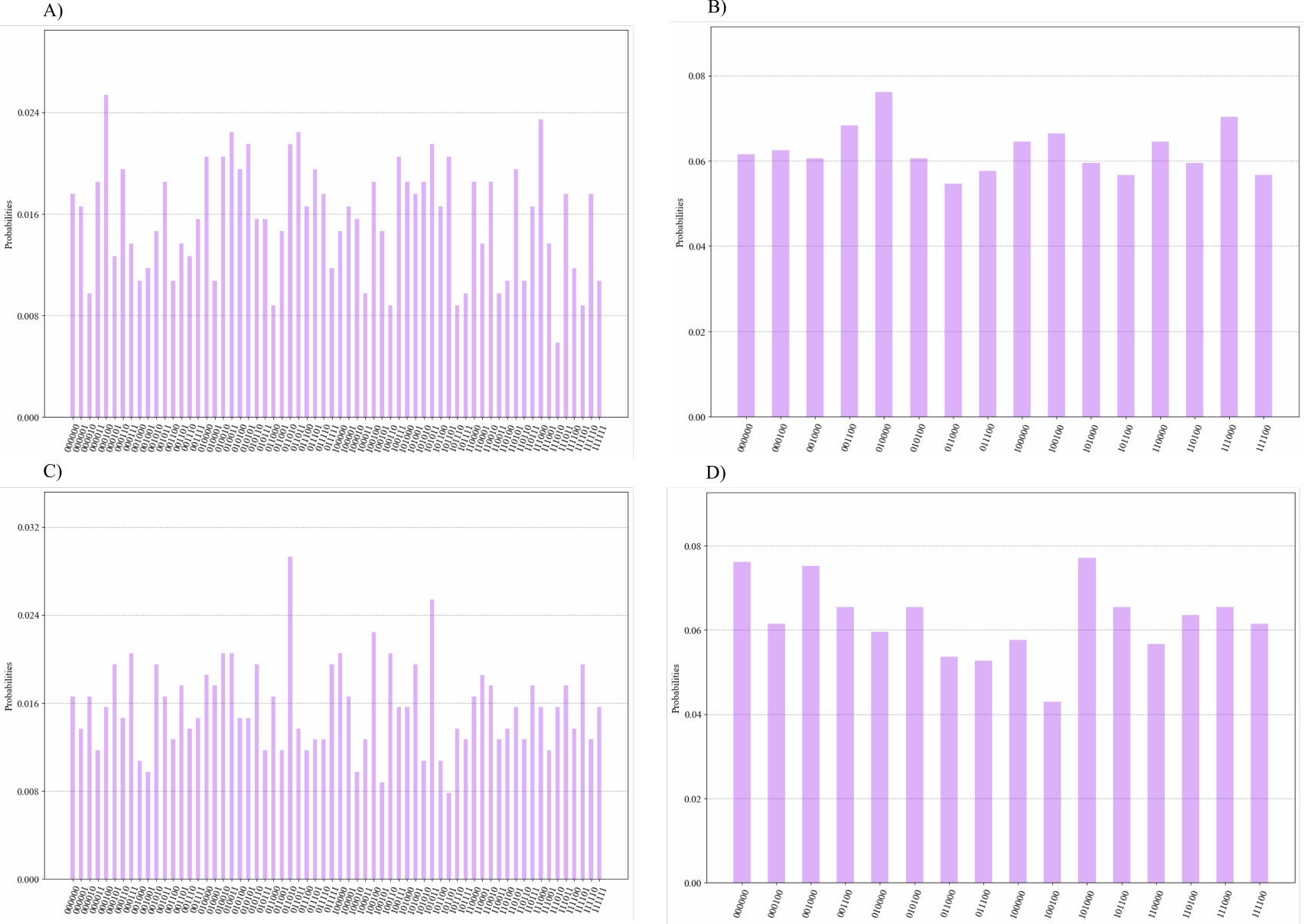}
  \caption{\label{fig:grover_clubs_variation_results}Results from performing the clubs search with four different Grover iteration counts.  Subfigure a uses two iterations, subfigure b uses four, subfigure c uses eight, and subfigure b uses ten.  The oscillatory behavior is clear: with four and ten iterations, the sixteen bitstrings representing club cards are correctly measured, but with two and eight iterations, \textit{all} bitstrings (including those representing non-club cards) are measured, and no result is measured more than 3\% of the time.}
  \Description[Simulation Results for Grover's Search Deck of Cards Example]{Bar graphs illustrating how too many iterations generated inaccurate results when simulating searching a deck of cards using Grover's Search Algorithm and Qiskit's simulators.  These indicate that when both too few (2) and too many (8) iterations are used, the results are not filtered as they should be; instead of measuring just 16 of the possible bitstrings, all 64 are measured in subfigures A and C.}
\end{figure}

We are now equipped to understand the oscillatory behavior we observed when simulating the deck of card searches described in Section \ref{sub:MustangQ_oracles}.
It is known that the ideal number of iterations for a Grover search is on the order of $\sqrt{N/M}$, where $N$ is the number of items in the database, and $M$ is the number of items for which we are searching \cite{nielsen11}.
However, because this characterization is a Big-O classification, for small databases such as the deck of cards examples, using the recommended $\sqrt{N/M}$ may generate poor results.
For example, for the clubs search, $\sqrt{N/M}=\sqrt{64/16}=2$, which is $O(10^0)$, as is $1, 3, 4, 5, 6, 7, 8,$ and $9$.
With two Grover iterations, the results were extremely poor (see subfigure a of Figure \ref{fig:grover_clubs_variation_results}).
Increasing the number of iterations from the `recommended' two to, say, eight produced results that were again very poor quality.  (See subfigure c of Figure \ref{fig:grover_clubs_variation_results}.)
Through further experimentation, we found the oscillating quality illustrated throughout Figure \ref{fig:grover_clubs_variation_results}, and verified that it occurs for other situations---such as the ten of diamonds search and Qiskit provided examples \cite{ibmq23}.
The explanation above can explain the `rapid' change in result quality; with just a few additional iterations, results change from very accurate to entirely inaccurate.
This is again a feature of the small size of the deck of cards database.
For this search, the initial $\theta/2$ is $\pi/6$, which means that each iteration moves the state by $\pi/3$ radians.
That is a significant change, meaning the state can loop back upon itself in just a few iterations. 
Consequently, for small databases, one must experimentally verify that the chosen number of iterations produces expected results, instead of solely relying upon the $O(\sqrt{N/M})$ recommendation.

\section{Appendix: Data for Section \ref{sub:MustangQ_QROMs}}\label{app:QROMs_data}

\begin{table}[hb]
\caption{Data for benchmark circuit \texttt{squar5}.\label{squar5}}
\begin{tabular}{cccccccccc}
\toprule
Encoding & Optimized & Inputs & Outputs & Qubits & \multicolumn{1}{p{1.5cm}}{\centering Natural \\ Gate Count} & \multicolumn{1}{p{1.5cm}}{\centering Natural \\ Complexity} & \multicolumn{1}{p{1.5cm}}{\centering Uniform \\ Gate Count} & \multicolumn{1}{p{1.5cm}}{\centering Uniform \\ Complexity} & \multicolumn{1}{p{1.5cm}}{\centering Uniform \\ Quantum Depth} \\
\midrule
Basis & No & 5 & 8 & 17 & 152 & 373 & 1800 & 2433 & 1092 \\
Basis & Yes & 5 & 8 & 17 & 134 & 355 & 1782 & 2415 & 1092 \\
Angle & No & 5 & 8 & 6 & 172 & 322 & 14992 & 17752 & 12092 \\ 
Angle & Yes & 5 & 8 & 6 & 64 & 96 & 512 & 576 & 448 \\
Improved Angle & No & 5 & 8 & 6 & 207 & 402 & 19473 & 23061 & 15719 \\
Improved Angle & Yes & 5 & 8 & 6 & 128 & 192 & 576 & 672 & 512 \\
\bottomrule
\end{tabular}
\end{table}

\begin{table}
\caption{Data for benchmark circuit \texttt{Z9sym}.\label{Z9sym}}
\begin{tabular}{cccccccccc}
\toprule
Encoding & Optimized & Inputs & Outputs & Qubits & \multicolumn{1}{p{1.5cm}}{\centering Natural \\ Gate Count} & \multicolumn{1}{p{1.5cm}}{\centering Natural \\ Complexity} & \multicolumn{1}{p{1.5cm}}{\centering Uniform \\ Gate Count} & \multicolumn{1}{p{1.5cm}}{\centering Uniform \\ Complexity} & \multicolumn{1}{p{1.5cm}}{\centering Uniform \\ Quantum Depth} \\
\midrule
Basis & No & 9 & 1 & 18 & 1065 & 2407 & 11401 & 15327 & 7257 \\
Basis & Yes & 9 & 1 & 18 & 803 & 2145 & 11139 & 15065 & 7257 \\
Angle & No & 9 & 1 & 10 & 4200 & 7980 & 3437280 & 4080720 & 2790061 \\ 
Angle & Yes & 9 & 1 & 10 & 768 & 1280 & 7936 & 8960 & 6912 \\
Improved Angle & No & 9 & 1 & 10 & 8400 & 15960 & 6874560 & 8161440 & 5580121 \\
Improved Angle & Yes & 9 & 1 & 10 & 1536 & 2560 & 8704 & 10240 & 7680 \\
\bottomrule
\end{tabular}
\end{table}

\begin{table}
\caption{Data for benchmark circuit \texttt{inc}.\label{inc}}
\begin{tabular}{cccccccccc}
\toprule
Encoding & Optimized & Inputs & Outputs & Qubits & \multicolumn{1}{p{1.5cm}}{\centering Natural \\ Gate Count} & \multicolumn{1}{p{1.5cm}}{\centering Natural \\ Complexity} & \multicolumn{1}{p{1.5cm}}{\centering Uniform \\ Gate Count} & \multicolumn{1}{p{1.5cm}}{\centering Uniform \\ Complexity} & \multicolumn{1}{p{1.5cm}}{\centering Uniform \\ Quantum Depth} \\
\midrule
Basis & No & 7 & 9 & 22 & 702 & 1787 & 8814 & 11927 & 5782 \\
Basis & Yes & 7 & 9 & 22 & 600 & 1685 & 8712 & 11825 & 5774 \\
Angle & No & 7 & 9 & 8 & 896 & 1624 & 212016 & 251536 & 171706 \\ 
Angle & Yes & 7 & 9 & 8 & 256 & 384 & 2048 & 2304 & 1792 \\
Improved Angle & No & 7 & 9 & 8 & 972 & 1756 & 228332 & 270892 & 184914 \\
Improved Angle & Yes & 7 & 9 & 8 & 702 & 1787 & 8814 & 11927 & 1960 \\
\bottomrule
\end{tabular}
\end{table}

\begin{table}
\caption{Data for benchmark circuit \texttt{Z5xp1}.\label{Z5xp1}}
\begin{tabular}{cccccccccc}
\toprule
Encoding & Optimized & Inputs & Outputs & Qubits & \multicolumn{1}{p{1.5cm}}{\centering Natural \\ Gate Count} & \multicolumn{1}{p{1.5cm}}{\centering Natural \\ Complexity} & \multicolumn{1}{p{1.5cm}}{\centering Uniform \\ Gate Count} & \multicolumn{1}{p{1.5cm}}{\centering Uniform \\ Complexity} & \multicolumn{1}{p{1.5cm}}{\centering Uniform \\ Quantum Depth} \\
\midrule
Basis & No & 7 & 10 & 23 & 412 & 981 & 4524 & 6121 & 2897 \\
Basis & Yes & 7 & 10 & 23 & 351 & 921 & 4464 & 6061 & 2895 \\
Angle & No & 7 & 10 & 8 & 1024 & 1920 & 260864 & 309504 & 211330 \\ 
Angle & Yes & 7 & 10 & 8 & 256 & 384 & 2048 & 2304 & 1792 \\
Improved Angle & No & 7 & 10 & 8 & 1157 & 2228 & 311747 & 369887 & 252605 \\
Improved Angle & Yes & 7 & 10 & 8 & 512 & 768 & 2304 & 2688 & 2048 \\
\bottomrule
\end{tabular}
\end{table}

\begin{table}
\caption{Data for benchmark circuit \texttt{dist}.\label{dist}}
\begin{tabular}{cccccccccc}
\toprule
Encoding & Optimized & Inputs & Outputs & Qubits & \multicolumn{1}{p{1.5cm}}{\centering Natural \\ Gate Count} & \multicolumn{1}{p{1.5cm}}{\centering Natural \\ Complexity} & \multicolumn{1}{p{1.5cm}}{\centering Uniform \\ Gate Count} & \multicolumn{1}{p{1.5cm}}{\centering Uniform \\ Complexity} & \multicolumn{1}{p{1.5cm}}{\centering Uniform \\ Quantum Depth} \\
\midrule
Basis & No & 8 & 5 & 20 & 1562 & 3952 & 19770 & 26712 & 12789 \\
Basis & Yes & 8 & 5 & 20 & 1348 & 3738 & 19556 & 26498 & 12781 \\
Angle & No & 8 & 5 & 9 & 2287 & 4327 & 1042177 & 1236997 & 845326 \\ 
Angle & Yes & 8 & 5 & 9 & 512 & 768 & 4096 & 4608 & 3584 \\
Improved Angle & No & 8 & 5 & 9 & 2588 & 5052 & 1258612 & 1493924 & 1021021 \\
Improved Angle & Yes & 8 & 5 & 9 & 1024 & 1536 & 4608 & 5376 & 4096 \\
\bottomrule
\end{tabular}
\end{table}

\begin{table}
\caption{Data for benchmark circuit \texttt{f51m}.\label{f51m}}
\begin{tabular}{cccccccccc}
\toprule
Encoding & Optimized & Inputs & Outputs & Qubits & \multicolumn{1}{p{1.5cm}}{\centering Natural \\ Gate Count} & \multicolumn{1}{p{1.5cm}}{\centering Natural \\ Complexity} & \multicolumn{1}{p{1.5cm}}{\centering Uniform \\ Gate Count} & \multicolumn{1}{p{1.5cm}}{\centering Uniform \\ Complexity} & \multicolumn{1}{p{1.5cm}}{\centering Uniform \\ Quantum Depth} \\
\midrule
Basis & No & 8 & 8 & 22 & 352 & 797 & 3504 & 4737 & 2249 \\
Basis & Yes & 8 & 8 & 22 & 284 & 729 & 3436 & 4669 & 2245 \\
Angle & No & 8 & 8 & 9 & 2295 & 4335 & 1042185 & 1237005 & 845327 \\ 
Angle & Yes & 8 & 8 & 9 & 512 & 768 & 4096 & 4608 & 3584 \\
Improved Angle & No & 8 & 8 & 9 & 3421 & 6485 & 1565295 & 1857907 & 1269647 \\
Improved Angle & Yes & 8 & 8 & 9 & 896 & 1408 & 4480 & 5248 & 3968 \\
\bottomrule
\end{tabular}
\end{table}

\begin{table}
\caption{Data for benchmark circuit \texttt{mlp4}.\label{mlp4}}
\begin{tabular}{cccccccccc}
\toprule
Encoding & Optimized & Inputs & Outputs & Qubits & \multicolumn{1}{p{1.5cm}}{\centering Natural \\ Gate Count} & \multicolumn{1}{p{1.5cm}}{\centering Natural \\ Complexity} & \multicolumn{1}{p{1.5cm}}{\centering Uniform \\ Gate Count} & \multicolumn{1}{p{1.5cm}}{\centering Uniform \\ Complexity} & \multicolumn{1}{p{1.5cm}}{\centering Uniform \\ Quantum Depth} \\
\midrule
Basis & No & 8 & 8 & 23 & 949 & 2494 & 12693 & 5196 & 8206 \\
Basis & Yes & 8 & 8 & 23 & 865 & 2410 & 12609 & 17090 & 8206 \\
Angle & No & 8 & 8 & 9 & 1905 & 3705 & 919455 & 1091355 & 745876 \\ 
Angle & Yes & 8 & 8 & 9 & 512 & 768 & 1050119 & 1246467 & 851956 \\
Improved Angle & No & 8 & 8 & 9 & 2073 & 4129 & 1050119 & 1246467 & 851956 \\
Improved Angle & Yes & 8 & 8 & 9 & 844 & 1356 & 4428 & 5196 & 3916 \\
\bottomrule
\end{tabular}
\end{table}

\begin{table}
\caption{Data for benchmark circuit \texttt{clip}.\label{clip}}
\begin{tabular}{cccccccccc}
\toprule
Encoding & Optimized & Inputs & Outputs & Qubits & \multicolumn{1}{p{1.5cm}}{\centering Natural \\ Gate Count} & \multicolumn{1}{p{1.5cm}}{\centering Natural \\ Complexity} & \multicolumn{1}{p{1.5cm}}{\centering Uniform \\ Gate Count} & \multicolumn{1}{p{1.5cm}}{\centering Uniform \\ Complexity} & \multicolumn{1}{p{1.5cm}}{\centering Uniform \\ Quantum Depth} \\
\midrule
Basis & No & 9 & 5 & 22 & 1379 & 3242 & 15635 & 21062 & 10011 \\
Basis & Yes & 9 & 5 & 22 & 1099 & 2962 & 15355 & 20782 & 1011 \\
Angle & No & 9 & 5 & 10 & 4956 & 9420 & 4059260 & 4819132 & 3294929 \\ 
Angle & Yes & 9 & 5 & 10 & 1024 & 1536 & 8192 & 9216 & 7168 \\
Improved Angle & No & 9 & 5 & 10 & 9093 & 17382 & 7537347 & 8948319 & 6118204 \\
Improved Angle & Yes & 9 & 5 & 10 & 2048 & 3072 & 9216 & 10752 & 8192 \\
\bottomrule
\end{tabular}
\end{table}

\begin{table}
\caption{Data for benchmark circuit \texttt{addm4}.\label{add_m4}}
\begin{tabular}{cccccccccc}
\toprule
Encoding & Optimized & Inputs & Outputs & Qubits & \multicolumn{1}{p{1.5cm}}{\centering Natural \\ Gate Count} & \multicolumn{1}{p{1.5cm}}{\centering Natural \\ Complexity} & \multicolumn{1}{p{1.5cm}}{\centering Uniform \\ Gate Count} & \multicolumn{1}{p{1.5cm}}{\centering Uniform \\ Complexity} & \multicolumn{1}{p{1.5cm}}{\centering Uniform \\ Quantum Depth} \\
\midrule
Basis & No & 9 & 8 & 25 & 3683 & 8922 & 43875 & 59162 & 28278 \\
Basis & Yes & 9 & 8 & 25 & 3005 & 8244 & 43197 & 58484 & 28278 \\
Angle & No & 9 & 8 & 10 & 4702 & 9022 & 3928222 & 4663582 & 3188641 \\
Angle & Yes & 9 & 8 & 10 & 1024 & 1536 & 8192 & 9216 & 7168 \\
Improved Angle & No & 9 & 8 & 10 & 4731 & 9078 & 3952773 & 4692729 & 3208570 \\
Improved Angle & Yes & 9 & 8 & 10 & 2042 & 3066 & 9210 & 10746 & 8186 \\
\bottomrule
\end{tabular}
\end{table}

\begin{table}
\caption{Data for benchmark circuit \texttt{b11}.\label{b11}}
\begin{tabular}{cccccccccc}
\toprule
Encoding & Optimized & Inputs & Outputs & Qubits & \multicolumn{1}{p{1.5cm}}{\centering Natural \\ Gate Count} & \multicolumn{1}{p{1.5cm}}{\centering Natural \\ Complexity} & \multicolumn{1}{p{1.5cm}}{\centering Uniform \\ Gate Count} & \multicolumn{1}{p{1.5cm}}{\centering Uniform \\ Complexity} & \multicolumn{1}{p{1.5cm}}{\centering Uniform \\ Quantum Depth} \\
\midrule
Basis & No & 8 & 31 & 44 & 416 & 1031 & 4848 & 6571 & 3126 \\
Basis & Yes & 8 & 31 & 44 & 364 & 979 & 4796 & 6519 & 3122 \\
Angle & No & 8 & 31 & 9 & 2304 & 4352 & 1046272 & 1241856 & 848642 \\ 
Angle & Yes & 8 & 31 & 9 & 320 & 576 & 3904 & 4416 & 3392 \\
Improved Angle & No & 8 & 31 & 9 & 2304 & 4352 & 1046272 & 1241856 & 848642 \\
Improved Angle & Yes & 8 & 31 & 9 & 576 & 1088 & 4160 & 4928 & 3648 \\
\bottomrule
\end{tabular}
\end{table}

\begin{table}
\caption{Data for benchmark circuit \texttt{apex4}.\label{apex4}}
\begin{tabular}{cccccccccc}
\toprule
Encoding & Optimized & Inputs & Outputs & Qubits & \multicolumn{1}{p{1.5cm}}{\centering Natural \\ Gate Count} & \multicolumn{1}{p{1.5cm}}{\centering Natural \\ Complexity} & \multicolumn{1}{p{1.5cm}}{\centering Uniform \\ Gate Count} & \multicolumn{1}{p{1.5cm}}{\centering Uniform \\ Complexity} & \multicolumn{1}{p{1.5cm}}{\centering Uniform \\ Quantum Depth} \\
\midrule
Basis & No & 9 & 19 & 36 & 61556 & 172859 & 907908 & 1230799 & 598257 \\
Basis & Yes & 9 & 19 & 36 & 58784 & 170087 & 905136 & 1228027 & 598255 \\
Angle & No & 9 & 19 & 10 & 5120 & 9728 & 4190208 & 4974592 & 3401218 \\
Angle & Yes & 9 & 19 & 10 & 1024 & 1536 & 8192 & 9216 & 7168 \\
Improved Angle & No & 9 & 19 & 10 & 5120 & 9728 & 4190208 & 4974592 & 3401218 \\
Improved Angle & Yes & 9 & 19 & 10 & 2560 & 1536 & 8704 & 10240 & 7680 \\
\bottomrule
\end{tabular}
\end{table}

\begin{table}
\caption{Data for benchmark circuit \texttt{ex5}.\label{ex5}}
\begin{tabular}{cccccccccc}
\toprule
Encoding & Optimized & Inputs & Outputs & Qubits & \multicolumn{1}{p{1.5cm}}{\centering Natural \\ Gate Count} & \multicolumn{1}{p{1.5cm}}{\centering Natural \\ Complexity} & \multicolumn{1}{p{1.5cm}}{\centering Uniform \\ Gate Count} & \multicolumn{1}{p{1.5cm}}{\centering Uniform \\ Complexity} & \multicolumn{1}{p{1.5cm}}{\centering Uniform \\ Quantum Depth} \\
\midrule
Basis & No & 8 & 63 & 78 & 7024 & 19525 & 101920 & 138145 & 67128 \\
Basis & Yes & 8 & 63 & 78 & 6678 & 19179 & 101574 & 137799 & 67128 \\
Angle & No & 8 & 63 & 9 & 2304 & 4352 & 1046272 & 1241856 & 848642 \\ 
Angle & Yes & 8 & 63 & 9 & 512 & 768 & 4096 & 4608 & 3584 \\
Improved Angle & No & 8 & 63 & 9 & 2688 & 4992 & 1177152 & 1397184 & 954722 \\
Improved Angle & Yes & 8 & 63 & 9 & 776 & 1288 & 4360 & 5128 & 3848 \\
\bottomrule
\end{tabular}
\end{table}

\clearpage

\section{Appendix: Data for Section \ref{sub:MustangQ_QRNGs}}\label{app:qrng_data}

\begin{table}[ht!]
\caption{Selecting a number of shots that reduces sampling error when simulating circuits generating each distribution.\label{table:qrng_shot_selection}}
\begin{tabular}{cccccc}
\toprule
Distribution & G-Statistic & Similarity Metric & Shots($\times$1.5) \\
\midrule
 Uniform & 0.0012 & 0.9723 & 34500 \\
 \hline
 Binomial & 0.0013 & 0.9712 &  21000 \\ 
 \hline
 Triangle & 0.001 & 0.9747 & 34500 \\ 
 \hline
 Bimodal Non-parametric & 0.001 & 0.9747 & 6000 \\ 
 \hline
 Arbitrary Non-parametric & 0.0008 & 0.9774 &  42000 \\
\bottomrule
\end{tabular}
\end{table}

\begin{table}[ht!]
\caption{Statistical comparison of distributions generated by QRNG subcircuits and the theoretical distributions they were designed to represent when simulated using IBM's Washington device noise signature.\label{table:qrng_noisy_simulation}}
\begin{tabular}{ccccc}
\toprule
Distribution & G-statistic & Similarity Metric & Gate Count & Quantum Depth \\
\midrule
Uniform & 0.0025 & 0.9601224 & 5 & 1 \\
Binomial & 1.9761 & 0.1598019 & 61 & 57 \\ 
Triangle & 0.0683 & 0.7938283 & 61 & 57 \\ 
Bimodal Non-Parametric & 0.0227 & 0.8802398 & 61 & 57 \\ 
Arbitrary Non-Parametric & 0.1031 & 0.7481408 & 61 & 57 \\ 
\bottomrule
\end{tabular}
\end{table}

\begin{table}[ht!]
\caption{Statistical comparison of distributions generated by optimized QRNG subcircuits and the theoretical distributions they were designed to represent when simulated using IBM's Washington device noise signature.\label{table:qrng_symmetric_simulation}}
\begin{tabular}{ccccc}
\toprule
Distribution & Similarity Metric & Gate Count & Quantum Depth \\
\midrule
Binomial & 0.3229 & 36 & 30 \\ 
Triangle & 0.9614 & 36 & 30 \\ 
Bimodal Non-Parametric & 0.9274 & 36 & 30 \\
\bottomrule
\end{tabular}
\end{table}

\clearpage

\section{Appendix: Data for Section \ref{sub:MustangQ_oracles}}\label{app:oracle_data}

\begin{table}[ht!]
\caption{Results for a set of benchmark functions with ESOP synthesis (and with no RTT method preprocessing).} 
\label{table:esop_no_expansion_results}
\begin{tabular}{cccccccc}
\toprule
Function & Inputs & Outputs & Qubits & Gate Count & Complexity & Time-to-Synthesis ($\mu$s) \\
\midrule
squar5	& 5 & 8 & 13 & 52 & 150 & 3.20e4 \\
Z9sym & 9 & 1 & 10 & 157 & 530 & 4.95e4  \\
inc & 7 & 9 & 16 & 118 & 441 & 3.24e4 \\
Z5xp1 & 7 & 10 & 17 & 100 & 291 & 4.16e4 \\
dist & 8 & 5 & 13 & 220 & 918 & 9.80e4 \\
f51m & 8 & 8 & 16 & 88 & 239 & 3.74e4 \\
mlp4 & 8 & 8 & 16 & 147 & 615 & 6.30e4 \\
clip & 9 & 5 & 14 & 188 & 824 & 6.09e4 \\
addm4 & 9 & 8 & 17 & 219 & 942 & 1.34e5 \\
b11 & 8 & 31 & 39 & 132 & 517 & 5.14e4 \\
apex4 & 9 & 19 & 28 & 5565 & 35393 & 2.38e7 \\
ex5 & 8 & 63 & 71 & 756 & 4374 & 2.49e5 \\
\bottomrule
\end{tabular}
\end{table}

\begin{table}[ht!]
\caption{Results for a set of benchmark functions with ESOP synthesis and RTT method preprocessing. Asterisks (*) denote a function that failed to synthesize due to too many inputs for explicit \texttt{.pla} expansion.} 
\label{table:esop_expansion_results}    
\begin{tabular}{cccccccc}
\toprule
Function & Inputs & Outputs & Qubits & Gate Count & Complexity & Time-to-Synthesis ($\mu$s) \\
\midrule
squar5	& 5 & 8 & 18 & 176 & 778 & 7.18e6 \\
Z9sym & 9 & 1 & 20 & 1255 & 8431 & 8.18e6 \\
inc & 7 & 9 & 28 & 773 & 6936 & 1.55e7 \\
Z5xp1 & 7 & 10 & 20 & 587 & 3150 & 2.63e6  \\
dist & 8 & 5 & 20 & 1189 & 7186 & 6.99e6 \\
f51m & 8 & 8 & 16 & 88 & 239 & 3.25e6 \\
mlp4 & 8 & 8 & 26 & 1092 & 8613 & 1.15e7 \\
clip & 9 & 5 & 22 & 1306 & 8645 & 8.08e6 \\
addm4 & 9 & 8 & 26 & 2480 & 18856 & 1.77e6 \\
b11 & 8 & 31 & * & * & * & * \\
apex4 & 9 & 19 & * & * & * & *  \\
ex5 & 8 & 63 & * & * & * & * \\
\bottomrule
\end{tabular}
\end{table}

\begin{table}[ht!]
\caption{Results for a set of benchmark functions with TBS, which requires fully-specified bijective \texttt{.pla} representations that we obtain via RTT method preprocessing. Asterisks (*) denote a function that failed to synthesize due to either too many inputs for explicit \texttt{.pla} expansion or too many required qubits/gates for TBS (more than 50,000 gates).}
\label{table:tbs_results} 
\begin{tabular}{cccccccc}
\toprule
Function & Inputs & Outputs & Qubits & Gate Count & Complexity & Time-to-Synthesis ($\mu$s) \\
\midrule
squar5	& 5 & 8 & 9 & 1463 & 3714 & 2.15e9 \\
Z9sym & 9 & 1 & 10 & 6028 & 17589 & 2.15e9 \\
inc & 7 & 9 & * & * & * & * \\
Z5xp1 & 7 & 10 & 10 & 5983 & 14924 & 2.15e9 \\
dist & 8 & 5 & 10 & 5737 & 14845 & 2.15e9 \\
f51m & 8 & 8 & 8 & 426 & 992 & 2.15e9 \\
mlp4 & 8 & 8 & * & * & * & *\\
clip & 9 & 5 & 11 & 15396 & 42127 & 2.16e9 \\
addm4 & 9 & 8 & * & * & * & * \\
b11 & 8 & 31 & * & * & * & * \\
apex4 & 9 & 19 & * & * & * & * \\
ex5 & 8 & 63 & * & * & * & * \\
\bottomrule
\end{tabular}
\end{table}

\end{document}